\documentclass{aa}  
\usepackage{graphicx,xspace}
\usepackage{array}
\usepackage{txfonts}
\usepackage{natbib}
\usepackage{booktabs}
\usepackage[flushleft]{threeparttable}
\usepackage{lscape}
\usepackage{subcaption}
\usepackage{hyperref}
\usepackage{comment}
\bibpunct{(}{)}{;}{a}{}{,}

\title{Detection of an Unidentified Soft X-ray Emission Feature in NGC~5548}
\author{Liyi Gu \inst{1,2,3}, Junjie Mao \inst{4,5,6}, Jelle S. Kaastra \inst{1,3}, Missagh Mehdipour \inst{7}, 
Ciro Pinto \inst{8,9}, Sam Grafton-Waters \inst{10}, Stefano Bianchi \inst{11}, Hermine Landt \inst{12},  
Graziella Branduardi-Raymont \inst{13}, Elisa Costantini \inst{1}, Jacobo Ebrero \inst{14}, Pierre-Olivier 
Petrucci \inst{15}, Ehud Behar \inst{16}, Laura di Gesu \inst{17}, Barbara De Marco \inst{18}, Giorgio Matt \inst{11},  
Jake A. J. Mitchell \inst{12}, Uria Peretz \inst{16}, Francesco Ursini \inst{11}, and Martin Ward \inst{12} }
\date{May 2022}
\institute{
SRON Netherlands Institute for Space Research, Niels Bohrweg 4, 2333 CA Leiden, the Netherlands
\and
RIKEN High Energy Astrophysics Laboratory, 2-1 Hirosawa, Wako, Saitama 351-0198, Japan
\and 
Leiden Observatory, Leiden University, PO Box 9513, 2300 RA Leiden, The Netherlands
\and 
Department of Astronomy, Tsinghua University, Beijing 100084, China
\and 
Department of Physics, Hiroshima University, 1-3-1 Kagamiyama, Higashi Hiroshima, Hiroshima 739-8526, Japan
\and 
Department of Physics, University of Strathclyde, Glasgow G4 0NG, UK
\and 
Space Telescope Science Institute, 3700 San Martin Drive, Baltimore, MD 21218, USA
\and 
INAF – IASF Palermo, Via U. La Malfa 153, I-90146 Palermo, Italy
\and 
ESTEC/ESA, Keplerlaan 1, NL-2201AZ Noordwijk, the Netherlands
\and 
Mullard Space Science Laboratory, University College London, Holmbury St. Mary, Dorking, Surrey, RH5 6NT, UK
\and 
Dipartimento di Matematica e Fisica, Università degli Studi Roma Tre, via della Vasca Navale 84, 00146 Roma, Italy
\and 
Centre for Extragalactic Astronomy, Department of Physics, Durham University, South Road, Durham DH1 3LE, UK
\and 
Mullard Space Science Laboratory, University College London, Holmbury St Mary, Dorking, Surrey, RH5 6NT, UK
\and 
Telespazio UK for the European Space Agency (ESA), European Space Astronomy Centre (ESAC), Camino Bajo del Castillo, s/n,
28692 Villanueva de la Cañada, Madrid, Spain
\and 
Univ. Grenoble Alpes, CNRS, IPAG, 38000 Grenoble, France
\and 
Department of Physics, Technion, Haifa, Israel
\and 
Italian Space Agency (ASI), Via del Politecnico snc, 00133, Roma, Italy
\and 
Departament de Física, EEBE, Universitat Politècnica de Catalunya, Av. Eduard Maristany 16, 08019 Barcelona, Spain
}
\abstract{
    NGC~5548 is an X-ray bright Seyfert 1 active galaxy. It exhibits a variety of spectroscopic features in the soft X-ray band, including in particular the absorption by the AGN outflows of a broad range of ionization states, with column densities up to 10$^{27}$ m$^{-2}$, and having speeds up to several thousand kilometers per second. The known emission features are in broad agreement with photoionized X-ray narrow and broad emission line models. 
}{
    We report on an X-ray spectroscopic study using 1.1~Ms {\it XMM-Newton} and 0.9~Ms {\it Chandra} grating observations of NGC~5548 spanning two decades. The aim is to search and characterize any potential spectroscopic features in addition to the known primary spectral components that are already modeled in high precision.
}{
    For each observation, we model the data using a global fit including an intrinsic spectral energy distribution of the AGN and the known distant X-ray absorbers and emitters. We utilize as much knowledge from previous studies. The fit residuals are stacked and scanned for possible secondary features.
}{
    We detect a weak unidentified excess emission feature at $\sim 18.4$~{\AA} (18.1~{\AA} in the restframe). The feature is seen at $>5\sigma$ statistical significance taking into account the look elsewhere effect. No known instrumental issues, atomic transitions, and astrophysical effects can explain this excess. The observed intensity of the possible feature seems to anti-correlate in time with the hardness ratio of the source. However, the variability might not be intrinsic, it might be caused by the time-variable obscuration by the outflows. An intriguing possibility is the line emission from charge exchange between a partially ionized outflow and a neutral layer in the same outflow, or in the close environment. Other possibilities, such as emission from a highly-ionized component with high outflowing speed, cannot be fully ruled out. 
}
{}

\titlerunning{Unidentified feature in NGC~5548}
\authorrunning{L. Gu}

\begin{document}

\maketitle

\section{Introduction}

High resolution X-ray spectra of active galactic nuclei (AGNs) provide a powerful tool to study the physical condition
of matter in the proximity of the supermassive black hole \citep{turner2009, reynolds2016}. These materials are thought to be photoionized 
by the strong radiation field from the inner central engine (e.g., \citealt{kaastra2012}). Complicated absorption lines and edges 
have been discovered for more than 50\% of the Seyfert 1 galaxies \citep{crenshaw2003, mckernan2007, long2010}, suggesting line-of-sight outflow velocities of 100-1000 km s$^{-1}$ and column densities of $10^{24}-10^{27}$ m$^{-2}$, otherwise known as ``warm absorbers''. Photoionized gas components outside
the line-of-sight are found to be responsible for broad and narrow emission lines, as well as narrow radiative recombination 
continuum observed in soft X-rays \citep{kaastra2000, costantini2007, guainazzi2007, whewell2015, mao2018, mao2019, gw2020, gw2021}. Obscuration of the soft X-ray continuum and extreme ultraviolet (EUV) emission by outflowing gas with high velocity ($> 1000$ km s$^{-1}$), relatively high column density ($> 10^{26}$ m$^{-2}$), and low ionization states are recently reported for a few AGNs \citep{kaastra2014}. Furthermore, resonance absorption lines in the Fe-K band with higher velocity shifts have been detected in several radio-quiet sources, indicating outflows at quasi-relativistic velocity $\sim 0.1 c$ \citep{tombesi2010, parker2017, pinto2018, kosec2018b, reeves2020}.  

Apart from the above spectroscopic features, there have been claims of various secondary components in the AGN spectra. Despite being relatively uncertain, these possible weak features might add new insights to the physical picture of active galaxies. For instance, \citet{pounds2018} reported detection of redshifted absorption lines of ionized Fe, Ca, Ar, S, and Si with the {\it XMM-Newton} spectra of PG1211+143, suggesting inflow of matter with velocity of $0.3 c$ onto the black hole. A similar weak feature was reported by \citet{giustini2017} with the spectra of NGC 2617. Using the {\it XMM-Newton} reflection grating spectrometer (RGS) data of 1H0707-495, \citet{blustin2009} claimed weak broad emission lines from C, N, O, and Fe showing both redshifted and blueshifted wings, which were interpreted as a part of the reflection line emission component from the accretion disk. In a subsequent work, \citet{kosec2018b} showed that the blueshifted component becomes more significant as new observations were added, while the redshifted part is less certain as the significance remains low with the new data. There are also hints of weak transitions from highly excited states of \ion{N}{VII} and \ion{S}{XV} detected with X-ray and UV spectroscopy, which might originate from charge exchange between the ionized AGN outflows and the neutral environmental materials \citep{gu2017, mao2018}. Since many of these discoveries are made at the limits of the available data, the uncertainties of the claimed line detection are often difficult to be fully addressed \citep{vaughan2008}.

The archetypal Seyfert 1 galaxy NGC~5548 is one of the most extensively studied AGNs \citep{kaastra2000, bottorff2000, kaastra2014, mehd2015, cappi2016, goad2016, deh2019, deh22019, landt2015, landt2019, kriss2019, wildy2021}. It is arguably one of 
best active galaxies for the weak feature search, because (1) the available {\it XMM-Newton} and {\it Chandra} grating data of this object have accumulated to $>2$ Ms in total, making it one of the deepest spectroscopic AGN dataset so far; and (2) the primary spectral components have been modeled to good precision in previous works. It was the first target in which narrow X-ray absorption lines from warm absorbers were discovered \citep{kaastra2000}. These absorbers are continuously studied for the spectral and temporal properties 
\citep{steenbrugge2005, digesu2015, ebrero2016}. The deep multi-wavelength campaign of NGC 5548 during the obscuration phase in 
2013$-$2014 provided unprecedented constraints to the ionization states, column densities, and kinematics of the obscurers 
\citep{kaastra2014, mehd2015}. The heavily obscured state further offers a unique opportunity to accurately model the narrow emission lines and radiative recombination continua which stand out at a low continuum level \citep{steenbrugge2005, detmers2009, whewell2015, mao2018}.

\begin{table*}[!htbp]
    \centering
    \caption{NGC~5548 archival observation log}    
    \begin{threeparttable}
    \begin{tabular}{lccccc}
       \hline
       \hline
        Observatory & Grating & ID & Start Time & Exposure & Group$^{a}$ \\
                    &         &    &            & (ks)     &       \\
         \hline
        {\it Chandra} & LETGS & 330 & 1999-12-11 22:52:24 & 85.1 & C99 \\
        {\it Chandra} & LETGS & 3045 & 2002-01-18 15:58:06 & 168.9 & C02 \\
        {\it Chandra} & LETGS & 3383 & 2002-01-21 07:35:00 & 170.3 & C02 \\
        {\it Chandra} & HETGS & 3046 & 2002-01-16 06:13:37 & 152.0 & C02 \\        
        {\it Chandra} & LETGS & 5598 & 2005-04-15 05:19:22 & 115.9 & C05 \\
        {\it Chandra} & LETGS & 6268 & 2005-04-18 00:32:16 & 25.0  & C05 \\
        {\it Chandra} & LETGS & 7722 & 2007-08-14 20:59:02 & 98.6  & C07 \\
        {\it Chandra} & LETGS & 8600 & 2007-08-17 03:55:56 & 36.8  & C07 \\   
        {\it Chandra} & LETGS & 16369 & 2013-09-01 00:02:48 & 29.7  & C13 \\
        {\it Chandra} & LETGS & 16368 & 2013-09-02 10:34:19 & 67.4  & C13 \\        
        {\it Chandra} & LETGS & 16314 & 2013-09-10 08:18:59 & 121.9  & C13 \\        
        \hline
        {\it XMM-Newton} & RGS & 0089960301 & 2001-07-09 15:45:59 & 95.8 & X01 \\
        {\it XMM-Newton} & RGS & 0089960401 & 2001-07-12 07:34:56 & 39.1 & X01 \\
        {\it XMM-Newton} & RGS & 0720110301 & 2013-06-22 04:10:29 & 50.9 & X13s \\
        {\it XMM-Newton} & RGS & 0720110401 & 2013-06-29 23:50:30 & 57.0 & X13s \\
        {\it XMM-Newton} & RGS & 0720110501 & 2013-07-07 23:28:42 & 57.0 & X13s \\ 
        {\it XMM-Newton} & RGS & 0720110601 & 2013-07-11 23:11:43 & 57.0 & X13s \\  
        {\it XMM-Newton} & RGS & 0720110701 & 2013-07-15 22:56:29 & 57.0 & X13s \\   
        {\it XMM-Newton} & RGS & 0720110801 & 2013-07-19 22:40:42 & 58.0 & X13s \\   
        {\it XMM-Newton} & RGS & 0720110901 & 2013-07-21 22:32:18 & 57.0 & X13s \\   
        {\it XMM-Newton} & RGS & 0720111001 & 2013-07-23 22:24:17 & 57.0 & X13s \\   
        {\it XMM-Newton} & RGS & 0720111101 & 2013-07-25 22:15:00 & 57.0 & X13s \\  
        {\it XMM-Newton} & RGS & 0720111201 & 2013-07-27 22:06:35 & 57.0 & X13s \\  
        {\it XMM-Newton} & RGS & 0720111301 & 2013-07-29 21:58:06 & 57.0 & X13s \\  
        {\it XMM-Newton} & RGS & 0720111401 & 2013-07-31 21:49:48 & 57.0 & X13s \\  
        {\it XMM-Newton} & RGS & 0720111501 & 2013-12-20 14:01:39 & 57.0 & X13w \\  
        {\it XMM-Newton} & RGS & 0720111601 & 2014-02-04 09:33:43 & 57.0 & X14 \\  
        {\it XMM-Newton} & RGS & 0771000101 & 2016-01-14 05:52:27 & 37.0 & X16 \\
        {\it XMM-Newton} & RGS & 0771000201 & 2016-01-16 06:36:31 & 34.0 & X16 \\    
        {\it XMM-Newton} & RGS & 0861360101 & 2021-01-27 04:33:50 & 76.0 & X21 \\   
        \hline
        \hline
    \end{tabular}
    $^{a}$: C={\it Chandra}, X={\it XMM-Newton}, s=summer, w=winter.\\
    \end{threeparttable}
    \label{obs_log}
\end{table*}

In this work we perform a systematic search for weak spectral components in the NGC~5548 spectra in the soft X-ray band by re-analyzing all the available {\it XMM-Newton} RGS and {\it Chandra} low-energy transmission grating spectrometer (LETGS) data, based on the accumulated knowledge on the primary spectral components. The structure of the paper is as follows. Section 2 describes the data processing, spectral modeling, line detection, and the analysis of systematic uncertainties. In Section 3, we discuss the possible interpretation of the new emission feature detected. All errors quoted throughout the paper correspond to 68\% confidence level. The redshift of NGC~5548 is set to 0.017175 \citep{dv1991}.

\section{Analysis and results}

\subsection{Observations and data reduction}

We use a total of 10 {\it Chandra} and 19 {\it XMM-Newton} archival observations. For some observations, the data were taken in a (quasi-)consecutive period with the same instrumental setting. We combine such data together for better statistics. Table~\ref{obs_log} shows the observation log.

The {\it XMM-Newton} RGS data are used in combination with the European Photon Imaging Camera (EPIC) pn data. The RGS instruments were operated in the standard Spectro+Q mode, and the EPIC pn used the Small-Window mode with the thin filter. The data were processed using {\it XMM-Newton} Science Analysis System (SAS) v19.1. 
Periods of high flaring background, in which the particle background exceeds 0.4 count/s for pn, were filtered out for both instruments. The main RGS spectra used in this work were generated by stacking the RGS-1 and RGS-2 first-order spectra in adjacent observations (Table~\ref{obs_log}). The RGS data in the $7-35$~{\AA} range are used, and for pn, we use the $0.8-8.0$~keV range. The known gain problem in 2013 and 2014 with pn calibration causes poor fits near the energy of the gold M-edge of the telescope mirror, therefore the pn data between 2.0~keV and 2.4~keV are discarded from the fit. The spectra are identified as group X01, X13s, X13w, X14, X16, and X21, in which X13s is a combination of 12 {\it XMM-Newton} observations of summer 2013. The net RGS exposure is in total 1.1~Ms. 

For {\it Chandra}, we use all available LETGS/HRC-S data extracted from the TGCat archive \footnote{http://tgcat.mit.edu}. The multiple spectra of the same group (Table~\ref{obs_log}) and
the associated response files are combined using the CIAO $combine\_grating\_spectra$ tool. The data in the $5-70$~{\AA} range are used. For the 2002 observation, the high energy transmission grating spectrometer (HETGS) data are also available from TGCat. We use the $1.55-15.5$~{\AA} range for the high-energy grating (HEG) and $2.5-18$~{\AA} for the medium-energy grating (MEG). To correct the cross-calibration uncertainty, the MEG flux is scaled by a factor of 0.954 with respect to the HEG flux. After combining the observations in several adjacent periods, we come up with five grouped spectra, C99, C02, C05, C07, and C13. The archival LETGS observations sum up to 920~ks.

The optimal binning \citep{kaastra2016} was applied to all the spectra. The standard pipeline instrumental background has been improved using a Wiener filter, which smooth out the noisy features in the continuum \citep{gu2020}. The spectral modeling is done with SPEX version 3.06 \citep{kaastra1996,kaastra2020}. We use C-statistics for spectral fitting.

\begin{figure*}[!htbp]
\centering
\resizebox{1.0\hsize}{!}{\includegraphics[angle=0]{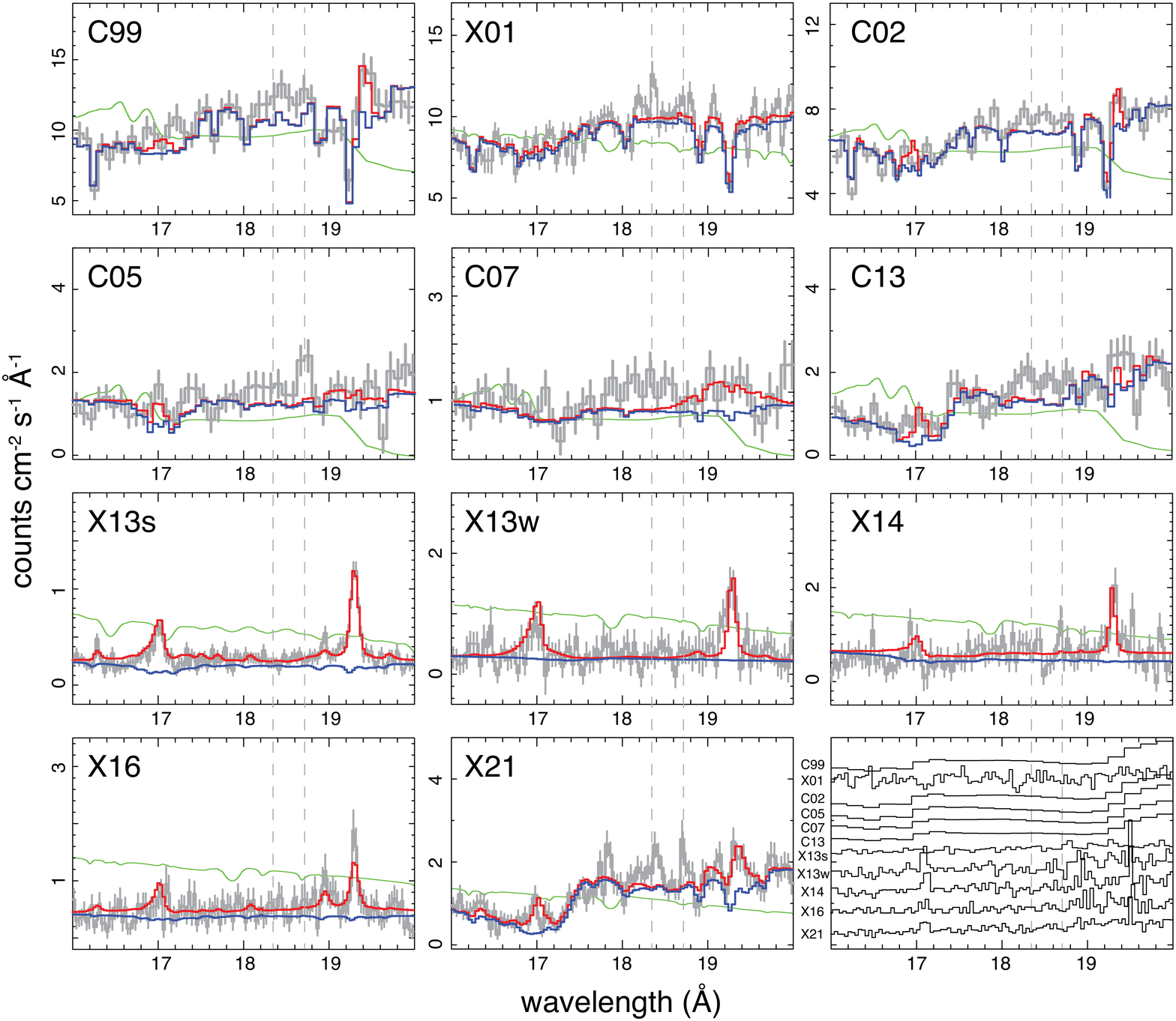}}
\caption{X-ray grating spectra of NGC~5548 in the 16$-$20~{\AA} band. All spectra are shown in the observed frame. The best-fit baseline models (described in \S~\ref{ana}) are shown in red. The baseline models excluding the emission {\it pion} components are plotted in blue. The thin green curves show the instrumental effective areas in arbitrary units. The corresponding background spectra are displayed in the bottom right panel. All background data are scaled for clarity. The vertical dashed lines mark the central wavelengths of the two possible peaks, 18.35~{\AA} and 18.72~{\AA}, seen in the X21 spectra (see \S~\ref{var} for detail).}
\label{fig:baseline}
\end{figure*}

\subsection{Analysis of the X-ray spectra}
\label{ana}

In order to model the known components including continuum, absorption, and emission at the same time, we analyze the {\it Chandra} and {\it XMM-Newton} spectra in the following way. The global model includes an intrinsic spectral energy distribution (SED), affected by the obscuration effect from multiple obscurers, the absorption from warm absorbers, as well as the absorption on the galactic scale. In addition, there are broad and narrow line features from the photoionized emitter. The baseline model is built upon the one used in \citet{mao2018} for the obscured 2013-2014 and 2016 spectra. For the early observations from 1999 to 2007, the obscuration components are not included as in \citet{ebrero2016} and in \citet{mao2017}. Basic components are summarized as follows.

The spectral energy distribution is described by the model consisting of a Comptonized soft X-ray excess ({\it comt}), a power law ({\it pow}) with exponential cut-off ($\it etau$) at high energy and low energy Lyman limit, and a disk reflection ($\it refl$) for modeling hard X-rays. The Comptonized component and the power law component are fed through obscurers and warm absorbers, modeled by two {\it pion} components and six {\it pion} components, respectively. The photoionization continuum received by the warm absorbers is the intrinsic SED affected by the obscuration. The outcome spectrum is corrected for a cosmological redshift $z = 0.017175$ and the Galactic absorption by a mainly neutral interstellar medium component using the {\it hot} model. The {\it hot} model has a fixed temperature of 0.5~eV, proto-solar abundances, and a hydrogen column density $N_{\rm H} = 1.45 \times 10^{24}$ m$^{-2}$ \citep{wakker2011}. 

The relevant parameters of the SED continuum are fixed to the values given in previous works (\citealt{ebrero2016} for C99, C02, C05, and C07; \citealt{mehd2015} for X01, X13s, and X16; and \citealt{ursini2015} for X13w and X14). The SED of C13 is set to the X13s values, while we allow the normalizations of the power law and reflection components to vary within the errors of the model in \citet{mehd2015}. As for the new data X21, the soft X-ray Comptonized component is fixed to the X16 one, and the normalization and $\Gamma$ of the dominant power law and the normalization of the reflection component are set free to fit. Results reveal rather minor differences between the X16 and X21 SEDs.

The column density $N_{\rm H}$, ionization parameter log $\xi$ and absorption covering factor $C_{\rm abs}$ of the obscuration components are set free in the fit. As for the warm absorbers, we fix their column density $N_{\rm H}$, bulk velocity $v_{\rm aver}$, and random motion $v_{\rm mic}$ to the values reported in \citet{kaastra2014, mao2017}, which came from a fit to the C02 data. We allow the ionization parameters log $\xi$ of the six warm absorbers to be different from previous values.

A time-average SED, including the {\it comt}, {\it pow}, and {\it refl} components are used to represent the ionizing continuum for the distant narrow line emission region. The narrow emission features are modeled with two {\it pion} components, in which the column density $N_{\rm H}$, ionization parameter log $\xi$, microscopic random motion $v_{\rm mic}$, average motion $v_{\rm aver}$, and emission scaling factor $C_{\rm em}$ are set as free parameters. Broad emission features are modeled with a third {\it pion} component, with $N_{\rm H}$, log $\xi$, and $C_{\rm em}$ free to fit. The absorption covering factors for the emission components are set to zero. Each component is convolved with a separate Gaussian velocity broadening component {\it vgau}, representing the macroscopic motion $v_{\rm mac}$ of the emitters.

The baseline fits are presented in Table~\ref{tab:param}. Overall, the fits reproduce well the spectra both during and outside the obscuration events. There are components in the present model that cannot be fully constrained with a few datasets due to the insufficient signal-to-noise level.
The warm absorber components for the C05 and C07 datasets are fixed to the values reported in \citet{ebrero2016}, and for X01, C07, C13, and X21, the current data can already be fit with two emission components instead of three. 

\onecolumn
\begin{landscape}
\begin{threeparttable}
\begin{longtable}{l|c@{\hskip 0.1in}c@{\hskip 0.1in}c@{\hskip 0.1in}c@{\hskip 0.1in}c@{\hskip 0.1in}c@{\hskip 0.1in}c@{\hskip 0.1in}c@{\hskip 0.1in}c@{\hskip 0.1in}c@{\hskip 0.1in}c}
\caption{Photoionized models of the absorption and emission components}\\
\hline
\hline
  & C99 & X01 & C02 & C05 & C07 & C13 & X13s & X13w & X14 & X16 & X21 \\
\hline
\endfirsthead
\caption{continued.}\\
\hline
\hline
  & C99 & X01 & C02 & C05 & C07 & C13 & X13s & X13w & X14 & X16 & X21 \\
\hline
\endhead
\hline
\endfoot
\multicolumn{12}{c}{obscuration component 1} \\
\hline
$N_{\rm H}$ (10$^{25}$ m$^{2}$) & $-$ & $-$ & $-$ & $-$ & $-$ & 11.7$\pm$0.2 & 12.4$\pm$0.1 & 10.7$\pm$0.3 & 11.4$\pm$0.3 & 15.4$\pm$0.5 & 11.3$\pm$0.4 \\
log($\xi$) (10$^{-9}$ W m) & $-$ & $-$ & $-$ & $-$ & $-$ & 0.71$\pm$0.05 & -0.08$\pm$0.04 & -4.0(f) & -1.97$\pm$0.32 & -1.57$\pm$0.62 & 0.97$\pm$0.04 \\
$C_{\rm abs}$ & $-$ & $-$ & $-$ & $-$ & $-$ & 0.95$\pm$0.05 & 0.94$\pm$0.02 & 0.84$\pm$0.03 & 0.96$\pm$0.04 & 0.94$\pm$0.06 & 0.98$\pm$0.01 \\
\hline
\multicolumn{12}{c}{obscuration component 2} \\
\hline
$N_{\rm H}$ (10$^{25}$ m$^{2}$) & $-$ & $-$ & $-$ & $-$ & $-$ & 89.5$\pm$8.7 & 94.3$\pm$4.4 & 101.2$\pm$9.1 & 91.2$\pm$8.8 & 78.3$\pm$7.4 & 34.4$\pm$1.4 \\
log($\xi$) (10$^{-9}$ W m) & $-$ & $-$ & $-$ & $-$ & $-$ & 0.004$\pm$0.1 & -3.2$\pm$1.1 & 1.6$\pm$0.1 & -0.2$\pm$0.1 & -3.5$\pm$0.5 & 2.1$\pm$0.3 \\
$C_{\rm abs}$ & $-$ & $-$ & $-$ & $-$ & $-$ & 0.42$\pm$0.12 & 0.42$\pm$0.06 & 0.64$\pm$0.07 & 0.20$\pm$0.05 & 0.22$\pm$0.04 & 0.42$\pm$0.03 \\
\hline
\multicolumn{12}{c}{warm absorbers} \\
\hline
log($\xi_1$) (10$^{-9}$ W m) & 0.59$\pm$0.10 & 0.90$\pm$0.12 & 0.37$\pm$0.07 & 0.07(f)$^{a}$ & 1.01(f) & 0.46$\pm$0.05 & 0.50$\pm$0.04 & -1.62$\pm$0.20 & -1.11$\pm$0.12 & 0.08$\pm$0.11 & -0.48$\pm$0.05   \\
log($\xi_2$) (10$^{-9}$ W m) & 1.93$\pm$0.12 & 1.24$\pm$0.07 & 1.40$\pm$0.10 & 0.01(f) & -2.08(f) & 1.37$\pm$0.11 & 0.86$\pm$0.05 & -0.02$\pm$0.03 & 1.35$\pm$0.05 & 1.30$\pm$0.17 & 0.02$\pm$0.06 \\
log($\xi_3$) (10$^{-9}$ W m) & 2.17$\pm$0.11 & 1.86$\pm$0.08 & 2.00$\pm$0.06 & 0.75(f) & 0.22(f) & 2.53$\pm$0.08 & 1.57$\pm$0.08 & 2.23$\pm$0.12 & 2.88$\pm$0.23 & 2.42$\pm$0.07 & 1.38$\pm$0.08 \\
log($\xi_4$) (10$^{-9}$ W m) & 2.20$\pm$0.13 & 1.89$\pm$0.10 & 1.93$\pm$0.05 & 1.80(f) & 1.98(f) & 2.25$\pm$0.09 & 2.38$\pm$0.07 & 1.89$\pm$0.10 & 1.92$\pm$0.11 & 1.72$\pm$0.45 & 2.89$\pm$0.15 \\
log($\xi_5$) (10$^{-9}$ W m) & 2.64$\pm$0.15 & 2.23$\pm$0.14 & 2.23$\pm$0.09 & 1.96(f) & 2.33(f) & 2.35$\pm$0.11 & 2.71$\pm$0.08 & 2.39$\pm$0.12 & 2.49$\pm$0.13 & 2.2$\pm$0.5 & 3.01$\pm$0.12 \\
log($\xi_6$) (10$^{-9}$ W m) & 2.42$\pm$0.20 & 2.70$\pm$0.15 & 2.72$\pm$0.10 & 2.59(f) & 2.30(f) & 4.03$\pm$0.21 & 2.74$\pm$0.09 & 2.63$\pm$0.09 & 2.67$\pm$0.21 & 3.5$\pm$1.5 & 3.23$\pm$0.16 \\
\hline
\multicolumn{12}{c}{X-ray narrow emission lines 1} \\
\hline
$N_{\rm H}$ (10$^{25}$ m$^{2}$) & 2.3$\pm$0.4 & 28.1$\pm$1.0 & 18.4$\pm$0.9 & 25.1$\pm$2.0 & 4.8$\pm$0.9 & 7.4$\pm$0.5 & 38.2$\pm$0.3 & 9.3$\pm$0.4 & 3.6$\pm$0.3 & 3.3$\pm$0.5 & 17.9$\pm$0.4 \\
log($\xi$) (10$^{-9}$ W m) & 0.61$\pm$0.05 & 0.62$\pm$0.03 & 0.82$\pm$0.04 & -0.16$\pm$0.03 & 0.68$\pm$0.2 & 0.70$\pm$0.06 & 0.27$\pm$0.03 & 0.67$\pm$0.04 & 0.46$\pm$0.05 & 0.46$\pm$0.04 & 0.86$\pm$0.05 \\
$C_{\rm em}$ & 0.10$\pm$0.02 & 0.01$\pm$0.002 & 0.01$\pm$0.003 & 0.01$\pm$0.01 & 0.03$\pm$0.01 & 0.04$\pm$0.01 & 0.01$\pm$0.01 & 0.03$\pm$0.01 & 0.04$\pm$0.01 & 0.07$\pm$0.02 & 0.02$\pm$0.01 \\
$v_{\rm mic}$ (km s$^{-1}$) & 0$\pm$60 & 700$\pm$80 & 50$\pm$50 & 0$\pm$100 & 0$\pm$120 & 0$\pm$70 & 0$\pm$30 & 440$\pm$120 & 10$\pm$210 & 10$\pm$280 & 90$\pm$80 \\
$v_{\rm aver}$ (km s$^{-1}$) & 48$\pm$20 & 240$\pm$40 & -440$\pm$40 & 1400$\pm$80 & -700$\pm$110 & 570$\pm$90 & -220$\pm$40 & -140$\pm$110 & -180$\pm$190 & -140$\pm$130 & -250$\pm$140 \\
\hline
\multicolumn{12}{c}{X-ray narrow emission lines 2} \\
\hline
$N_{\rm H}$ (10$^{25}$ m$^{2}$) & 23.8$\pm$1.2 & $-$ & 2.7$\pm$0.3 & 90.9$\pm$3.2 & $-$ & $-$ & 8.0$\pm$0.4 & 11.8$\pm$0.8 & 7.3$\pm$0.6 & 1.7$\pm$0.2 & 3.4$\pm$0.3 \\
log($\xi$) (10$^{-9}$ W m) & 1.52$\pm$0.12 & $-$ & 1.79$\pm$0.06 & 0.51$\pm$0.08 & $-$ & $-$ & 1.31$\pm$0.07 & 1.34$\pm$0.12 & 1.33$\pm$0.10 & 1.23$\pm$0.07 & 1.38$\pm$0.05 \\
$C_{\rm em}$ & 0.02$\pm$0.01 & $-$ & 0.05$\pm$0.01 & 0.002$\pm$0.001 & $-$ & $-$ & 0.02$\pm$0.01 & 0.01$\pm$0.01 & 0.02$\pm$0.01 & 0.05$\pm$0.01 & 0.03$\pm$0.01 \\
$v_{\rm mic}$ (km s$^{-1}$) & 500$\pm$100 & $-$ & 0$\pm$70 & 0$\pm$110 & $-$ & $-$ & 0$\pm$50 & 520$\pm$70 & 270$\pm$120 & 490$\pm$90 & 740$\pm$110 \\
$v_{\rm aver}$ (km s$^{-1}$) & 1990$\pm$80 & $-$ & 1250$\pm$100 & -460$\pm$120 & $-$ & $-$ & 0$\pm$40 & -240$\pm$200 & -10$\pm$250 & 20$\pm$180 & 1060$\pm$110 \\
\hline
\multicolumn{12}{c}{X-ray broad emission line} \\
\hline
$N_{\rm H}$ (10$^{25}$ m$^{2}$) & 0.3$\pm$0.1 & 14.0$\pm$0.5 & 19.3$\pm$0.6 & 28.0$\pm$1.9 & 42.5$\pm$4.5 & 0.9$\pm$0.5 & 5.0$\pm$0.4 & 6.8$\pm$0.6 & 26.8$\pm$1.2 & 29.7$\pm$0.9 & $-$ \\
log($\xi$) (10$^{-9}$ W m) & -0.3$\pm$0.1 & 1.2$\pm$0.1 & -0.48$\pm$0.12 & 1.47$\pm$0.19 & -0.03$\pm$0.3 & 0.12$\pm$0.02 & 1.34$\pm$0.04 & 1.78$\pm$0.15 & 1.35$\pm$0.14 & 1.39$\pm$0.09 & \\
$C_{\rm em}$ & 0.13$\pm$0.02 & 0.04$\pm$0.01 & 0.06$\pm$0.01 & 0.01$\pm$0.01 & 0.01$\pm$0.01 & 0.01$\pm$0.01 & 0.05$\pm$0.01 & 0.04$\pm$0.01 & 0.05$\pm$0.01 & 0.02$\pm$0.01 & $-$ \\
$v_{\rm mac}$ (km s$^{-1}$) & 5700$\pm$200 & 9800$\pm$600 & 4500$\pm$700 & 5100$\pm$1100 & 3200$\pm$700 & 6900$\pm$400 & 9600$\pm$500 & 9700$\pm$800 & 10030$\pm$900 & 8600$\pm$1200 & $-$ \\
$v_{\rm aver}$ (km s$^{-1}$) & -1830$\pm$500 & 0$\pm$400 & -1900$\pm$500 & 40$\pm$200 & 3700$\pm$500 & -120$\pm$190 & 4000$\pm$900 & 7300$\pm$1400 & 5600$\pm$1100 & 0$\pm$300 & $-$ \\
\hline
C-stat & 1625.37 & 2144.38 & 3855.88 & 1550.82 & 1379.98 & 1650.43 & 2941.37 & 1711.40 & 1689.61 & 1554.50 & 2223.93 \\
Expected & 1363.79 & 1579.01 & 3123.64 & 1367.89 & 1366.87 & 1364.10 & 2130.41 & 1342.95 & 1336.58 & 1357.11 & 1897.85 \\
\hline
\hline
\label{tab:param}
\end{longtable}
\noindent $^{a}$: These log $\xi$ cannot be well constrained with the current data. They are fixed to the values from \citet{ebrero2016}.
\end{threeparttable}
\end{landscape}
\twocolumn

\begin{figure*}[!htbp]
\centering
\resizebox{1.0\hsize}{!}{\includegraphics[angle=0]{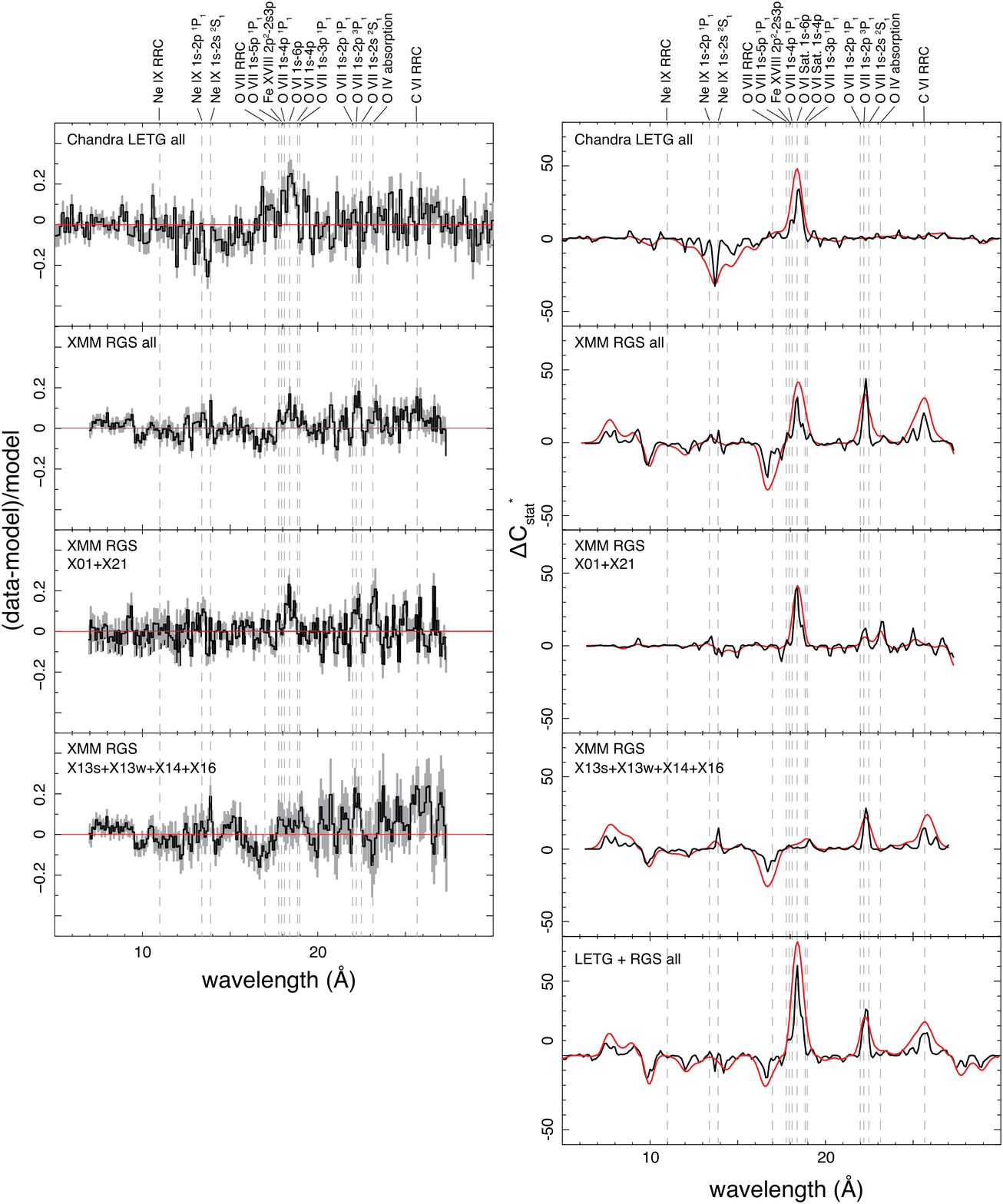}}
\caption{Stacked residuals with respect to the baseline fit (left), and Gaussian line scan performed on the data (right). In the right panel, the results for two different line widths $\sigma = 1000$ km s$^{-1}$ and 4000 km s$^{-1}$ are shown in black and red. Notable transitions relevant for the potential features in the stacked residuals are marked with vertical dashed lines. The wavelengths of the transitions are set to the AGN restframe.}
\label{fig:residual}
\end{figure*}

\subsection{Detection of a weak emission feature}
\label{detection}

After modeling the known emission and absorption components with the baseline model, we examine the fit residual for possible additional components. One unidentified residual feature is visually detected at a wavelength of $18.0-18.8$~{\AA} in the observed frame. As shown in Figure~\ref{fig:baseline}, the observed data exhibit weak excesses above the model that is used to fit the global spectra for C99, X01, C02, C05, C07, C13, and X21. For the other spectra, there is no to little evidence for the excess in $18.0-18.8$~{\AA}. Several spectra in particular X01, C02, C13, and X21 indicate that this feature might be composed of more than one narrow line-like peaks, while the others are more consistent with one peak. The main peak and the possible secondary peak are found around 18.4~{\AA} and 18.7~{\AA}. For clarity, hereafter we refer these peaks collectively as {\it one feature}. The X21 residual might also indicate a third weak peak at $\sim 17.8$~{\AA}, however, this peak is not seen in other spectra, furthermore, it overlaps in part with the known \ion{O}{VII} absorption lines at $\sim 17.6-17.7$~{\AA} and $17.9-18.0$~{\AA}. Therefore, we do not include the possible third peak in the subsequent discussion. C99 also might contain an excess at 17.2~{\AA}, but it is rather weak and not seen in other data. The line profile and the possible components of the detected feature will be discussed later in \S~\ref{var}.

To better constrain the possible excess feature, we stack the fit residuals of the {\it XMM-Newton} RGS and {\it Chandra} LETGS data. Following the method of \citet{gu2018}, individual residuals are combined with a weighting based on the counts in each energy bin. As shown in Figure~\ref{fig:residual}, several features can be identified in the stacked spectra. Both the LETGS and RGS residuals show clear peaks at $\sim 18.4$~{\AA} (restframe $18.1$~{\AA}), and the RGS residual might have additional peaks at 22.3~{\AA}, 23.1~{\AA}, and 25.6~{\AA}
which might coincide with the \ion{O}{VII} He$\alpha$ triplet, \ion{O}{VI} absorption line, and \ion{C}{VI} radiative recombination continuum. There are also
potential dips in the residuals, at the \ion{Ne}{IX} He$\alpha$ triplet ($\sim 13.6$~{\AA}) with the LETGS data and \ion{O}{VII} radiative recombination continuum ($\sim 17.0~{\AA}$) with the RGS data. Furthermore, by dividing the full RGS sample into two groups, X01 + X21 and the others, we see in Figure~\ref{fig:residual} that the $\sim 18.4$~{\AA} excess comes mostly from X01 + X21, while the other possible peaks and the 
dip seen in the full residual might originate from the other RGS data. The stacked LETGS residual at 18.4~{\AA} is contributed equally from multiple observations, instead of being dominated by a particular data. The possible temporal property of the excess is discussed in \S~\ref{var}.

The 18.4~{\AA} wavelength is not consistent with any strong emission or absorption lines in the baseline model. Indeed there are several known nearby atomic transitions that might be relevant, including \ion{O}{VI}, \ion{O}{VII}, and \ion{Fe}{XVIII} lines (Fig.~\ref{fig:residual}). The \ion{O}{VI} and \ion{O}{VII} transitions will be addressed later in \S~\ref{sys} and \S~\ref{cx}. The \ion{Fe}{XVIII} 2s 2p$^6$ $-$ 2s$^2$ 2p$^4$ 3p line is unlikely to be responsible for the excess emission,
since the same ion should also give transitions at 14.44~{\AA} and 16.34~{\AA} in the observed frame. The latter lines have much larger intensities than the former one for both collisional and photoionized plasmas, but they are not seen in the data.

Following the method of \citet{hitomiatomic}, we calculate the significance of the features by scanning the spectra with a Gaussian line. The Gaussian component has a central wavelength changing from 5~{\AA} to 30~{\AA}, and a width $\sigma$ setting to two trial values, 1000 km s$^{-1}$ and 4000 km s$^{-1}$. At each grid wavelength, we refit the spectra with baseline plus the Gaussian model, and calculate the $\Delta$C-stat improvement which is then multiplied by the sign of the best-fit normalization to distinguish emission and absorption features. As shown in Figure~\ref{fig:residual}, the 18.4~{\AA} excess is seen in the Gaussian line scan consistently for the stacked LETGS and the RGS data. The maximal $\Delta$C-stat improvements are 57 and 71 for the two kinds of line widths considered. This improvement comes from both instruments; the $\Delta$C-stat values with the $\sigma=1000$ km s$^{-1}$ scan are 35 and 45 for the stacked RGS and LETGS data, respectively.

The confidence level of the excess can be determined by taking into account the large amount of trials performed to find the line across the wavelength range (the so-called look-elsewhere effect). This effect is addressed in the appendix using both analytic and numerical approaches. Both results show that the putative 18.4~{\AA} feature can be detected above the confidence level of 5$\sigma$ after the look-elsewhere effect is accounted for. It is the only significant feature detected in both RGS and LETGS bands.

\subsection{Systematic effects}
\label{sys}

Here we investigate further the LETGS and RGS spectra to prove that the detected feature is not an instrumental artifact, a plasma code defect, or a known astrophysical effect. First we examine the effective area curves and instrumental background. As shown in Figure~\ref{fig:baseline}, there are small variations in particular for the RGS effective area at a $2-5$\% level, some of which are close to the position of the detected line feature. If these variations are not fully calibrated, they might induce artificial spectral features. 
However, the distributions of the variations are inconsistent with the excess at 18.4~{\AA}, and their amplitudes appear to be much weaker than the excess feature ($\sim 10-20$\%, Fig.~\ref{fig:residual}), therefore they cannot fully explain the detection.
It is even more unlikely for the line detected with the LETGS data to be a false feature in the effective area curves, which are fairly smooth in the wavelength range of interest. Similarly, the line feature cannot be due to known variations in the instrumental background spectra shown in Figure~\ref{fig:baseline}, because these variations are more than one order of magnitude weaker than the observed line intensity at 18.4~{\AA}. Furthermore, we do not see any other bright X-ray source in the RGS extraction regions that might potentially contaminate the observed spectra of the central object.

\begin{figure}[!htbp]
\centering
\resizebox{1.0\hsize}{!}{\includegraphics[angle=0]{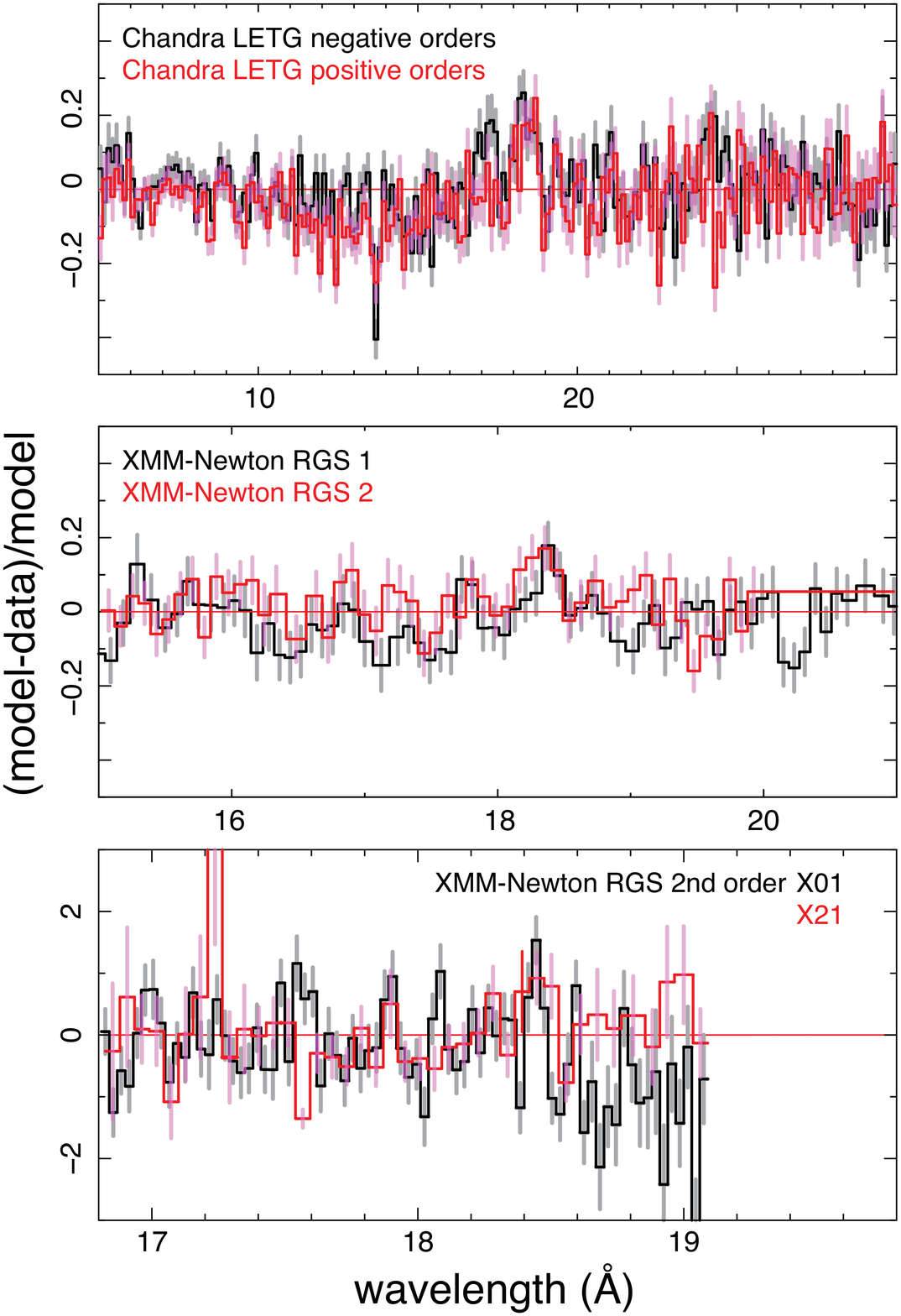}}
\caption{Comparison of the stacked fit residuals with the positive and negative orders of the LETGS (upper panel), the RGS 1 and RGS 2 instruments (middle). The residuals from the RGS second order data of the X01 and X21 observations are shown in the lower panel. }
\label{fig:order}
\end{figure}

To further address the instrumental effect, we fit separately the source spectra obtained in positive and negative orders of the LETGS, as well as with the RGS 1 and RGS 2 instruments. It can be seen from Figure~\ref{fig:order} that the stacked fit residuals with positive and negative LETGS orders overlap within their statistical uncertainties. Both data show an excess between 18~{\AA} and 19~{\AA}. Similarly, the residuals with the RGS 1 and RGS 2 agree upon an excess peaked at 18.4~{\AA}. In addition to that, we also show in Figure~\ref{fig:order} the residuals obtained with the second-order RGS spectra. A similar excess can be seen from the second order data but it is too noisy to provide useful constraints. The consistency between different orders and instruments of the grating data indicates that this excess line feature is unlikely to be an instrumental artifact.

\begin{figure}[!htbp]
\centering
\resizebox{1.0\hsize}{!}{\includegraphics[angle=0]{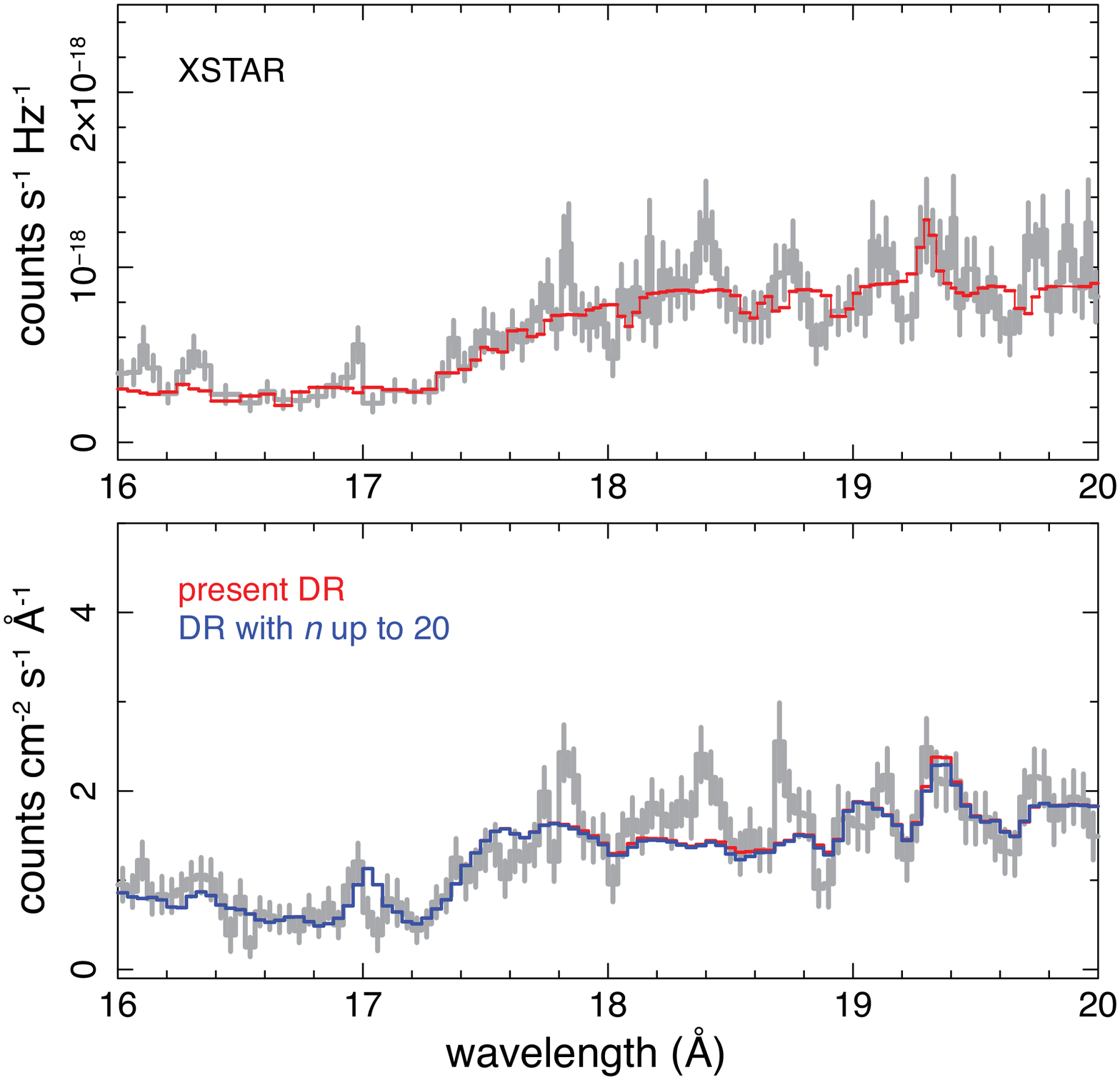}}
\caption{X21 spectrum and the corresponding fit with the XSTAR code (upper panel), and the baseline fit using SPEX with (blue) and without (red) highly excited $n > 5$ dielectronic recombination transitions (lower). }
\label{fig:xstar}
\end{figure}

One remaining possibility is that the feature is a false detection caused by an error in the spectral modeling, for instance, a missing transition in the plasma code at 18.4~{\AA}. Such a possibility might be examined by running the calculation with different plasma codes \citep{hitomiatomic}, and by fixing possible flaws in the present atomic database \citep{gu2019}. First we model the X21 spectrum using the XSTAR code. The spectral energy distribution of the ionizing source is set to be the same as in \S~\ref{ana}. The photoionization calculation is run with the semi-analytic {\it warmabs} model version 2.41 which is based on the standard population files pre-calculated with the XSTAR v2.58. The absorption and the emission {\it pion} components in the baseline are replaced by the absorbing {\it warmabs} and {\it photemis} models, respectively, while the other components are kept the same. The fit parameters of the XSTAR components are log $\xi$, $N_{\rm H}$, and velocity shift. As shown in Figure~\ref{fig:xstar}, there is no significant transition at 18.4~{\AA} with the best-fit XSTAR model, and the excess feature obtained with the SPEX and XSTAR models appears to be similar. 

We further test the spectral model by utilizing a more sophisticated atomic data calculation. The emission lines of H-like and He-like ions (in particular oxygen) are calculated up to high principal quantum number $n$ in SPEX, and none of them falls on 18.4~{\AA} (Fig.~\ref{fig:residual}). The Li-like satellite lines, which often show more complex pattern, are more plausible candidates of the excess emission. Several \ion{O}{VI} dielectronic recombination (DR) transitions indeed occur in the range $18.3-18.7$~{\AA}. These DR lines are calculated based on radiative cascade from doubly excited levels, while in the present atomic database we include doubly excited levels only up to $n=5$ for \ion{O}{VI}. To address the effect of the DR lines from higher shells, we put forward a new calculation of \ion{O}{VI} covering singly and doubly excited levels up to $n=20$. The level energies, transition probabilities, and radiative branching ratios are obtained in the same approach as described in \citet{gu2019}. We run the X21 fit again after implementing the new \ion{O}{VI} calculation in SPEX. As shown in Figure~\ref{fig:xstar}, the fit with new atomic data is nearly identical to the original one, indicating that the missing DR lines from high-$n$ levels in the photoionization model fall well below the level of the observed excess.

To quantify the excess over the known DR intensity, we allow the DR line emissivity free to fit. The most relevant transition
is 1s$^2$ 2s $-$ 1s 2s 6p at restframe wavelength 18.06~{\AA}. It can fully account for the observed excess in the X21 spectrum 
when the intensity of the transition increases by a factor of 62. This factor much exceeds the typical errors in atomic data of the DR lines \citep{gu2020}, as well as the possible line enhancement due to external electric field observed in ground-based laboratory \citep{bohm2003}. Furthermore, it would be rather unexpected that this line would be significantly underestimated, while its neighbour DR lines of similar origin are not, for a photoionized emission source.

\begin{figure}[!htbp]
\centering
\resizebox{1.0\hsize}{!}{\includegraphics[angle=0]{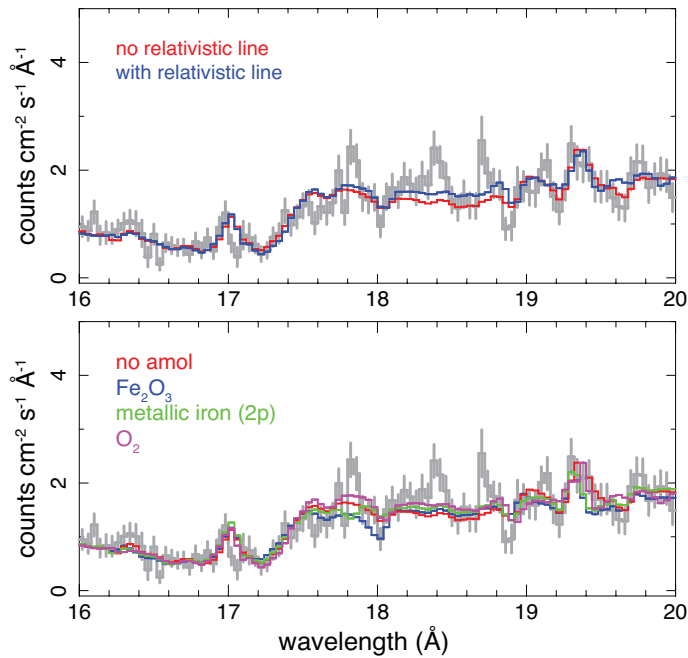}}
\caption{X21 spectrum and the fit with the baseline model broadened with a relativistic {\it laor} profile (upper panel), 
and the fit with the baseline plus an {\it amol} component (lower). Results with different compositions of the compound are plotted in blue, green, and magenta.}
\label{fig:laor}
\end{figure}

Another possible interpretation is a known astrophysical feature which is relevant to either a relativistic broad line, or a dust absorption component. As reported in \citet{br2001}, the RGS spectra of MCG-6-30-15 and Mrk 766 show significant excess emission between 18~{\AA} and 19~{\AA} above a non-relativistic warm absorber model. The observed spectra are better reproduced by relativistically broadened skewed Ly$\alpha$ lines of \ion{O}{VIII}, \ion{N}{VII}, and \ion{C}{VI}, originating from a combination of gravitational redshift around the black holes and the relativistic beaming due to gas swirling at extreme velocities. We examined this possibility on the NGC 5548 X21 spectrum, by adding an \ion{O}{VIII} Ly$\alpha$ line at rest-frame 18.96~{\AA} broadened with a relativistic {\it laor} profile \citep{laor1991} to the baseline model. The new best-fit model is plotted in Figure~\ref{fig:laor}. We find that the 
additional relativistic line component does not match with the observed excess at 18.4~{\AA} which is apparently narrower than the {\it laor} profile. Thus, the relativistic effect predicted from the standard model is unlikely the source of the feature. One potential caveat here is that the relativistic broadening profile obtained above could be a bit different from those derived by more sophisticated relativistic reflection models such as {\it relxill} \citep{garcia2014, dauser2022}, which calculate more accurately the angular dependence of the intrinsic reflection emission.

An alternative scenario for the excess features observed in the grating spectra of MCG-6-30-15 is the so called dusty warm absorbers \citep{julia2010}. It has been proposed that the excess features can be explained by the superposition of O VII absorption lines with the L-shell absorption complex of Fe I, which is likely caused by the dust potentially embedded in the partially ionized outflows \citep{julia2001}. To test this possibility for NGC~5548, we utilize an {\it amol} component in SPEX to model the absorption from the possible dusty or molecular matter around the warm absorbers. The most relevant {\it amol} compound is Fe$_2$O$_3$, which contains both O K- and Fe L- edges. At the same time, we also tested the cases with molecular O, as well as with metallic Fe components. As shown in Figure~\ref{fig:laor}, the dusty components do induce several broad features in the 16$-$20~{\AA} range, however, it cannot reproduce the relatively narrow feature at 18.4~{\AA}. Therefore, the present dust depleted outflow model does not explain the line.    

As a summary, we report a discovery of a weak emission feature at 18.4~{\AA} using the stacked grating spectrum of NGC~5548. The restframe wavelength is 18.1~{\AA}. The line is seen at $>5\sigma$ significance accounting for the look-elsewhere effect. We find that it is unlikely an instrumental feature, or a defect in the plasma model. Common astrophysical effects, including a relativistic broad line and a dust absorption component, cannot reproduce the observed excess.

\begin{figure*}[!htbp]
\centering
\resizebox{1.0\hsize}{!}{\includegraphics[angle=0]{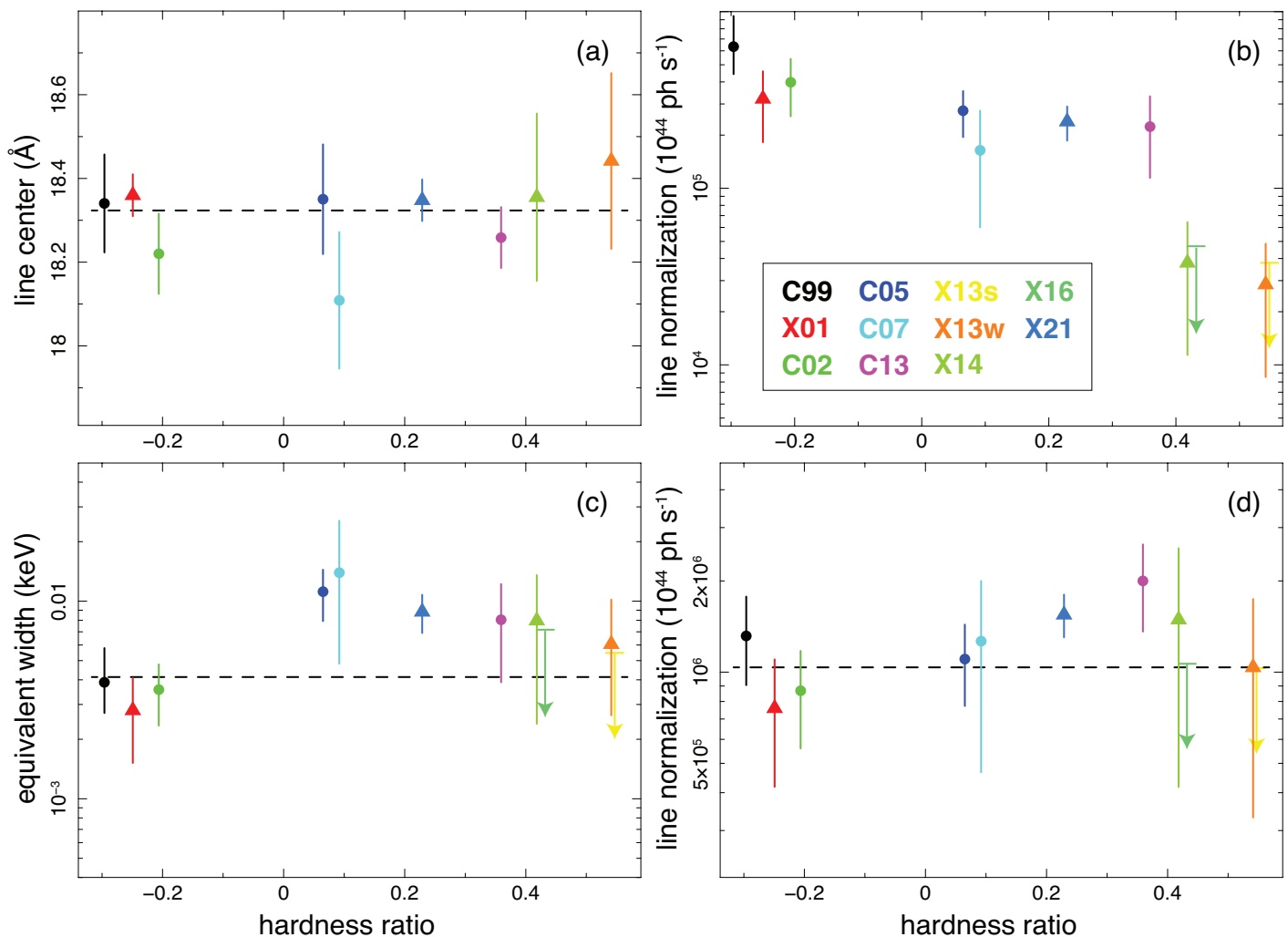}}
\caption{Variation of the observed excess feature as a function of the average hardness ratio $(H-S)/(H+S)$ (see \S~\ref{var}). Panels (a) $-$ (d) show the variations of the central wavelength of the G1 component described in \S~\ref{var}, the total G1 + G2 normalization assuming unabsorbed Gaussian components, the total equivalent width, and total normalization assuming absorption. The vertical lines in (a), (b), and (d) show the fits with a constant model.}
\label{fig:var}
\end{figure*}

\subsection{Line profile and variability}
\label{var}

The stacked grating spectra shown in Figure~\ref{fig:residual} illustrate an excess that appears to be broader than a single narrow line feature. As described in \S~\ref{detection}, several observations (Fig.~\ref{fig:baseline}) further imply that the excess feature might contain two emission peaks, one centered at around $18.1-18.5$~{\AA}, and the other at $18.4-18.8$~{\AA}. The best example, X21, shows clearly a double-peaked line profile at 18.35~{\AA} and 18.72~{\AA}. For all the observations except C05, the second peak at longer wavelength appears to be dimmer than the first one. To sufficiently model the observed excess, we added two Gaussian components into the baseline model. Their central wavelengths are free to vary within the ranges of $18.1-18.5$~{\AA} (hereafter G1) and $18.4-18.8$~{\AA} (hereafter G2), while the line intensities and widths are set free. In several cases the width of the second Gaussian component cannot be constrained, we fix it to that of the first one.      

This model allows us to assess the possible variation of the excess. First we assume that the Gaussian components remain unabsorbed by the outflowing components (i.e., X-ray obscurers and warm absorbers), while corrected only for the cosmological redshift and Galactic absorption using the {\it hot} component. The central wavelengths and the normalizations of the Gaussian components can be constrained for each observation. We plot in Figure~\ref{fig:var} the central wavelength variation of the G1 component as a function of the hardness ratio $(H-S)/(H+S)$ derived from the best-fit baseline model, where $H$ is the hard X-ray flux in the $1.5-10.0$~keV band and $S$ is the soft flux in $0.3-1.5$~keV. A large hardness ratio is often seen when strong X-ray obscuration occurs to this source \citep{mehd2015, mehd2017, kaastra2018, mehd2022}. We do not plot the wavelength of the G2 component as it is poorly constrained in several observations. It can be seen that the central wavelength of the G1 component does not vary significantly over the past 20 years, while the hardness ratio changes significantly between -0.3 and 0.5. Figure~\ref{fig:var}b plots the combined normalization of the two Gaussian components as a function of the hardness ratio. The line intensity appears to decrease as the hardness ratio increases. Fitting the normalization-hardness ratio diagram with a constant horizontal line yields a reduced $\chi^{2}$ of 10.05, indicating a probability p-value of 0.0015. It means that the intensity of the observed excess, if unabsorbed by the AGN absorbers, is significantly varying in the past observations. 

Interestingly, as shown in Figure~\ref{fig:var}c, the combined equivalent width of the two Gaussian components appears to be much less variable than the normalizations of the lines. This is because the obscuration plays an essential role in the continuum and the hardness ratio variation. As the obscuration increased around 2013 \citep{kaastra2014}, the continuum around 18.4~{\AA} decreased, which cancels out the decreasing intensity of the observed excess around that time, leading to a nearly time-constant equivalent width. This motivates us to address the second possibility: the excess component, represented by the two Gaussian lines, has been affected by the absorbing materials around the AGN, including the obscurers and the warm absorbers. As shown in Figure~\ref{fig:var}d, the combined line intensity from this modified model becomes nearly constant, which differs dramatically from Figure~\ref{fig:var}b where the line intensity occurs to be strongly variable as a function of the hardness ratio. Therefore the observed excess could in fact be a constant component if its emission  
is partially obscured by the outflowing clouds found at $\sim 0.01$ to several parsec from the central engine \citep{kaastra2014}.

\begin{figure}[!htbp]
\centering
\resizebox{1.0\hsize}{!}{\includegraphics[angle=0]{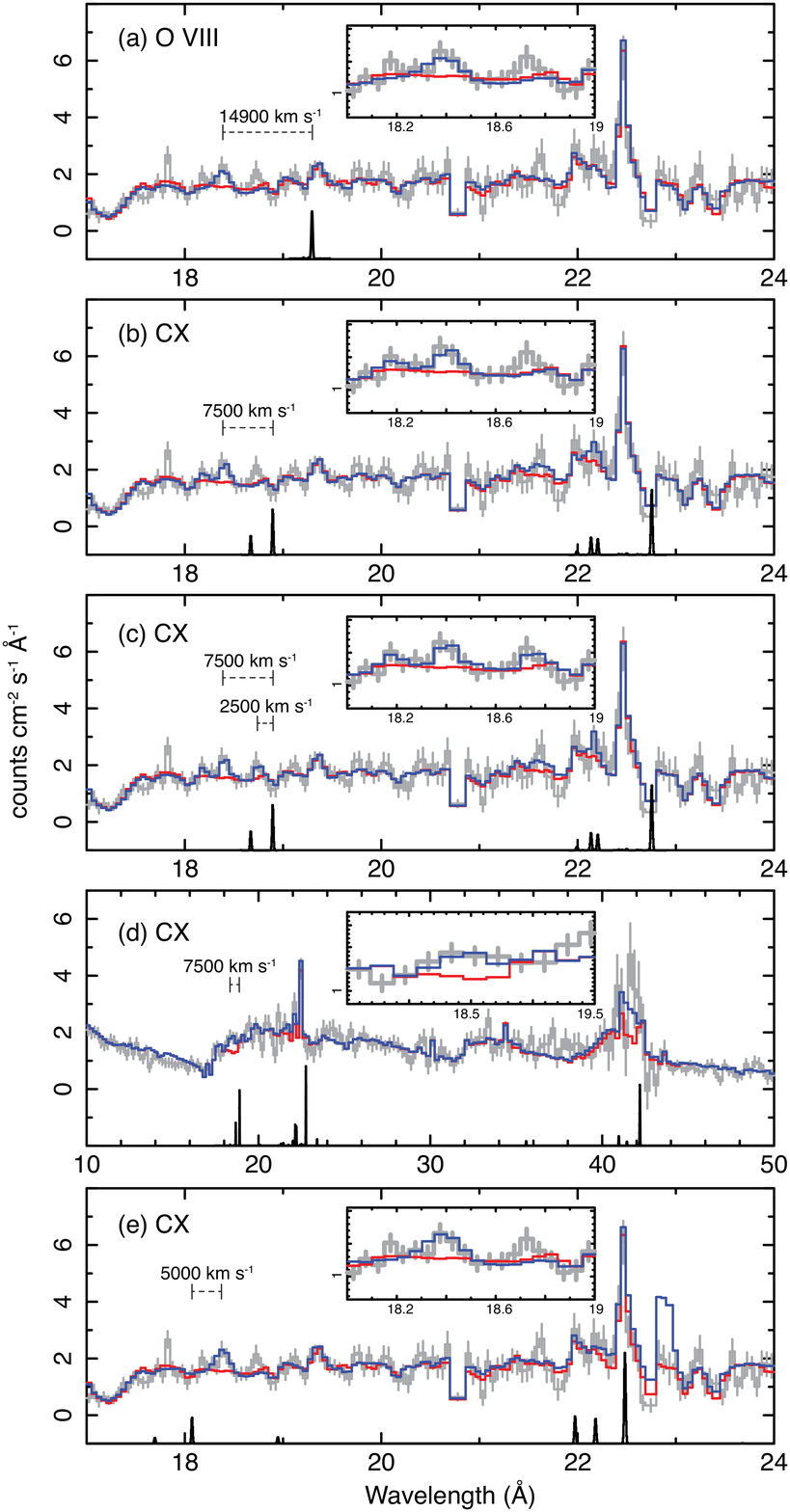}}
\caption{NGC~5548 spectra fit with the baseline plus an additional component. The X21 data and the best-fit blueshifted {\it pion} model with log($\xi$) $= 2$ (\S~\ref{oviii}) are shown in (a). Panels (b) and (c) plot the X21 spectrum fit with the blueshifted single charge exchange of ionization temperature = 0.05~keV, and two charge exchange components of the same temperature. The C13 data with single charge exchange component is shown in (d). The X21 spectrum with a redshifted charge exchange component (ionization temperature = 0.1~keV) is displayed in (e). In all panels, the baseline model is plotted in red, and the baseline plus additional component is shown in blue. The relevant lines of the additional components ($pion$ and $cx$) are also plotted below the data. The wavelengths of the transitions are given at the AGN restframe, and the intensities are scaled for clarity. For better comparison, the spectra and the models around the 18.4~{\AA} feature are also shown in the insets. }
\label{fig:dis}
\end{figure}

\section{Discussion}

By examining the stacked {\it XMM-Newton} RGS and {\it Chandra} LETGS spectra of the archetypal Seyfert 1 galaxy NGC 5548,
we detect a weak emission feature around 18.4~{\AA} which consists of one or two peaks above the model. There is no likely candidate for an atomic transition in a photoionized or collisional plasma, except for several very weak satellite transitions from $n \geq 4$ of \ion{O}{VI}. It is also unlikely to be
residuals due to absorption features from matter in gas or dust forms. This signal appears to be relatively bright in years 1999$-$2013 and 2021, but rather dim from 2013 to 2016 when the strong obscuration by outflows occurred to this source \citep{kaastra2014, mehd2015}. It is unclear whether this possible variation is intrinsic to the source, or it is caused by 
the attenuation from the X-ray obscuration and absorption near the center.

A firm interpretation of the observed excess is hindered by the fact that it is merely one or two weak lines. 
Below we discuss two candidate scenarios, one with a shifted emission line and the other incorporating a different plasma process.

\subsection{Blueshifted \ion{O}{VIII} line}
\label{oviii}

\citet{blustin2009} reported possible broad excess features found with the 
{\it XMM-Newton} RGS spectrum of Seyfert 1 galaxy 1H0707$-$495. In their Figure~2, a broad emission feature is seen
at restframe 18.6~{\AA}, which has been interpreted as a blueshifted component of \ion{O}{VIII} line. Deeper {\it XMM-Newton} data of 1H0707$-$495 reveal that the emitting gas has a radial velocity of about 8000 km s$^{-1}$, and 
a velocity dispersion of $3000-6000$ km s$^{-1}$ \citep{kosec2018b, xu2021}. Similar blueshifted emission features 
are found at the wavelengths of \ion{N}{VII}, \ion{Fe}{XVII}, \ion{Ne}{X}, \ion{S}{XVI}, and \ion{Fe}{XXV}/\ion{Fe}{XXVI}. 
These lines are interpreted as photoionized emission from a wind component powered by the radiation pressure due to high accretion rate of the AGN.

Assuming that the NGC~5548 feature is a shifted component of \ion{O}{VIII} Ly$\alpha$ line, the average Doppler velocity is 
obtained to be $14900 \pm 600$ km s$^{-1}$ and the average Gaussian broadening is $800 \pm 200$ km s$^{-1}$. The average speed is much higher while the random motion is milder than those of the blueshifted emission component in 1H0707$-$495. 
The excess emission in NGC~5548 would 
originate from a significantly ionized (e.g., log $\xi$ $\geq 2$) component, since no blueshifted \ion{O}{VII} counterpart can be 
detected in the stacked spectrum. As shown in Figure~\ref{fig:dis}, the observed excess in the X21 spectrum seems to be consistent with 
a blueshifted log $\xi$ = 2 {\it pion} emission component which is added to the baseline model. The new component does not predict any detectable features from other elements elsewhere in
the spectrum including the soft X-ray and Fe K-shell bands. We do not see any evidence for other possible relativistic emission or absorption line in the {\it Chandra} LETGS and the {\it XMM-Newton} RGS as well as the EPIC spectra.
It is therefore not feasible to constrain further the ionization state of the possible emitter with the present data. It should noted that the radial Doppler velocity obtained above is clearly higher than those of the known warm absorbers in X-ray 
\citep{kaastra2014,mehd2015,digesu2015,ebrero2016} as well as the kinematic structures seen at longer wavelengths \citep{crenshaw2003,shapovalova2004,arav2015,li2016}. It is in better agreement with those of the ultrafast outflows (e.g., \citealt{tombesi2013, parker2018}), although the ultrafast outflows are often observed as blueshifted absorption lines rather than an emission feature.

\subsection{Charge exchange model}
\label{cx}

Another potential scenario is that the detected excess comes from an extra plasma component. Here we address the 
possibility of a charge exchange line \citep{gu2015,gu2016,gu2017,gu2018}. The charge exchange emission originates from capture of a single electron from the target neutral hydrogen atom to a highly charged projectile ion that collides with the neutral target,
\begin{equation}
X^{q+} + H \longrightarrow X^{(q-1)+}(nl ^{2S+1}L) + H^{+} + \delta E,
\end{equation}
where $X^{q+}$ is the highly charged ion, $nl$ denotes the principal quantum number and orbital angular momentum of the captured electron, $^{2S+1}L$ is the total spin $S$ and total orbital angular momentum $L$, and $\delta E$ is the kinetic energy released in the collision. The electron is first captured into a highly excited state with large quantum number $n$, then relaxes via cascade.

As pointed out in \S~\ref{sys}, the best matching atomic transitions of the observed feature are the Li-like oxygen satellite lines from states with one core excitation electron (1s $-$ 2s), and the other electron at a shell with large principal quantum number $n$. Although most existing charge exchange calculations do not consider core excitation (e.g., \citealt{smith2012,smith2014,gu2016,cumbee2018}), this process does occur, and the core excitation with charge exchange has already been observed in laboratory for the Li-like line production (e.g., \citealt{pepmiller1983,tanis1985,lee1991} and references therein). Although oxygen was not tested in their experiments, it is natural to expect the same. The core excitation process in charge exchange is often referred as transfer excitation, which occurs simultaneously with the electron capture, forming doubly excited intermediate states. The excitation can be mediated either by the Coulomb field interaction of the target proton, or through the resonant ion-electron interaction analogously to dielectronic recombination. Laboratory measurements on Ne$^{6+}$ + He collision \citep{beijers1992}
suggested that the charge exchange cross section with core excitation can be comparable with those for the conventional core conserving capture
at low collision velocities. Other measurements further demonstrated that the captured electron will fall primarily onto high-$n$ orbits in both core excitation and core conserving cases (e.g., \citealt{raphaelian1991,tanis1985}). Therefore we consider the charge exchange with core excitation as a plausible candidate producing the doubly-excited high-$n$ Li-like O lines.

Because the charge exchange with core excitation is not available in the present SPEX {\it cx} model \citep{gu2016}, we build a simple model on O$^{6+}$ (\ion{O}{VII}) + H to test our scenario. This process would end up with a proton and a O$^{5+}$ (\ion{O}{VI}) ion in excited state. First we assume that the total cross section with core excitation is the same as the present core-conserving process that is available in SPEX. Then we let the core electron excite via 1s $-$ 2s and 1s $-$ 2p channels,
and apply the present velocity-dependent $n-$ and $l-$ distributions, obtained in the theoretical calculation using quantum molecular-orbital close-coupling method \citep{wu2012}, to the level-resolved cross sections of the captured electron. The new calculations are inserted in the SPEX atomic database, making it possible to evaluate the spectrum of O$^{6+}$ + H taking into account both the core exciting and core conserving processes. As the O$^{6+}$ (\ion{O}{VII}) has the strongest emission lines in the observed spectra, it is feasible to expect that this ion is responsible for the most relevant charge exchange feature.

This simple model predicts that the core conserving charge exchange of O$^{6+}$ + H produces \ion{O}{VI} emission lines (e.g., 2s $-$ 4p
and 2s $-$ 5p) mostly at restframe $\sim 110$~{\AA}, while the core exciting capture gives several transitions including in
particular the 1s$^2$ 2s $-$ 1s 2s 4p (18.35~{\AA} and 18.57~{\AA}) and 1s$^2$ 2s $-$ 1s 2s 2p (22.36~{\AA}), which might be
relevant to the observed excess in NGC~5548. By refitting the X21 spectrum with the baseline plus the new {\it cx} model, we
find that the primary excess feature at 18.4~{\AA} is consistent with the 1s$^2$ 2s $-$ 1s 2s 4p line blueshifted by a
Doppler velocity of 7500 $\pm 500$ km s$^{-1}$. The best-fit ionization temperature is 0.05 $\pm 0.02$ keV, and the velocity
dispersion of the charge exchange component is measured to be $\leq 1050$ km s$^{-1}$. Adding the new charge exchange component improves the C-statistics by
51 for an expected value of 1893, with a probability of $7\times10^{-9}$ that the better fit is caused by chance. As shown
in Figure~\ref{fig:dis}, besides the \ion{O}{VI} lines, the new component introduces a few \ion{C}{V} lines, including in
particular 1s$^2$ $-$ 1s 2s, 1s$^2$ $-$ 1s 2p, and 1s$^2$ $-$ 1s 3p at restframe 41.47~{\AA}, 40.27~{\AA}, and 34.97~{\AA}.
These \ion{C}{V} lines are consistent with the LETGS data, although the $\Delta$C-stat is not enough for a significant detection.
Any other charge exchange lines in the soft X-ray and the UV ranges are expected to be at least 50 weaker than the
\ion{O}{VI} 1s$^2$ 2s $-$ 1s 2s 4p line, they are unlikely to be visible with the existing observations using {\it Chandra}, {\it XMM-Newton}, and {\it Hubble}.

It is possible that these charge exchange lines come from mixing of a warm outflow, partially ionized by either photon or collision, with an adjacent cold layer that is not yet ionized. The cold matter could be a part of the outflow itself, or it may belong to a component in the close environment, e.g., the dusty torus. The outflow velocity of the possible O$^{6+}$ gas is slightly larger than that of the X-ray obscurer, $\leq$ 5000 km s$^{-1}$, measured from the associated broad UV absorption lines in the Hubble cosmic origins spectrograph data \citep{kaastra2014}. It could be compared with the velocity dispersions of the broad UV (up to $\sim 6800$ km s$^{-1}$, \citealt{kaastra2014}) and X-ray (7400 km s$^{-1}$, \citealt{mao2018}) emission. These agreements suggest that the observed emission feature might originate from the central region of the AGN. However, it must be noted that the current estimations of the radial velocity as well as the line intensities using the charge exchange model fully rely on the $nl$-distribution of core exciting O$^{6+}$ + H capture, which is obtained based on crude assumption. The line wavelengths and intensities should further vary as a function of the O$^{6+}$ + H collision velocity that is still very uncertain. Any further implication on the origin of the possible emission would be too speculative before the present charge exchange modeling is verified with an $\it ab$ $\it initio$ theoretical calculation or a dedicated laboratory measurement.

The possible secondary peak at $18.4-18.8$~{\AA} (\S~\ref{var}) seen in particular in X21 and C05 cannot be accounted for by the same charge exchange component. It might indicate another velocity component within the outflowing cloud. To test this possibility, we refit the X21 spectrum using the baseline plus two charge exchange components. All parameters of the second component, except the normalization, redshift, and velocity dispersion, are fixed to those of the first one. The possible peak at 18.7~{\AA} can be reproduced by a {\it cx} component blueshifted by an outflowing velocity of 2500 $\pm 900$ km s$^{-1}$, with a line width $\leq 400$ km s$^{-1}$. 

Besides O$^{6+}$ + H, the core conserving capture of O$^{7+} +$ H produces emission lines at restframe wavelengths of 17.77~{\AA} (1s$^2$ $-$ 1s 4p) and 17.40~{\AA} (1s$^2$ $-$ 1s 5p), providing another potential candidate for the observed excess. These lines are calculated based on the existing {\it cx} model which incorporates the cross section calculated using the quantum molecular-orbital close-coupling method (e.g., \citealt{nolte2012}). To fit the observed excess, the 1s$^2$ $-$ 1s 4p and 1s$^2$ $-$ 1s 5p lines need to be redshifted by Doppler velocities of $\sim 5000$ km s$^{-1}$ and $\sim 11000$ km s$^{-1}$ with respect to the AGN. In practice, however, the O$^{7+} +$ H charge exchange would also produce a strong redshifted 1s $-$ 2s forbidden line that was not seen in the observed data (Fig.~\ref{fig:dis}e). The 2s level is mostly populated by cascades from $n=4$ and 5 intermediate levels. Reducing the 1s $-$ 2s transition would require a major revision to the $l-$ and $S-$ distributions of the captured electron, which is rather unlikely given the available results from the laboratory measurements \citep{mullen2016, cumbee2018}. Therefore, the O$^{7+} +$ H charge exchange cannot be
the primary process.

\section{Conclusion}

By reanalyzing all the available {\it XMM-Newton} RGS and {\it Chandra} LETGS spectra of the Seyfert 1 galaxy NGC~5548, we detect a emission-like feature at 18.4~{\AA} (restframe 18.1~{\AA}). This feature is weak, with equivalent width $\leq 10$~eV. Despite of this, the stacked significance reaches 5$\sigma$ taking into account the look elsewhere effect. The new feature seems to be either intrinsically variable in intensity, or affected by the absorption components that changed over time. We demonstrate that the known systematic issues, including the calibration defects, the missing atomic lines, and the secondary astrophysical effects, cannot significantly affect the detection.  The remaining possibilities are the charge exchange emission from the outflowing wind of moderate radial velocity, or a photoionized emission component from the very high speed wind. Disentangling these possibilities is impossible given the present data, and has to wait until the launch of {\it XRISM}, {\it Arcus}, and {\it Athena}.

\begin{acknowledgements}

SRON is supported financially by NWO, the Netherlands Organization for
Scientific Research. SB acknowledges financial support from ASI under grants ASIINAFI/037/12/0 and n. 2017-14-H.O and from PRIN MIUR project “Black Hole winds and the Baryon Life Cycle of Galaxies: the stoneguest at the galaxy evolution supper", contract no. 2017PH3WAT. BDM acknowledges support from a Ramón y Cajal Fellowship (RYC2018-025950-I).
\end{acknowledgements}

\bibliographystyle{aa}
\bibliography{main}

\begin{thebibliography}{82}
\expandafter\ifx\csname natexlab\endcsname\relax\def\natexlab#1{#1}\fi

\bibitem[{{Arav} {et~al.}(2015){Arav}, {Chamberlain}, {Kriss}, {Kaastra},
  {Cappi}, {Mehdipour}, {Petrucci}, {Steenbrugge}, {Behar}, {Bianchi},
  {Boissay}, {Branduardi-Raymont}, {Costantini}, {Ely}, {Ebrero}, {di Gesu},
  {Harrison}, {Kaspi}, {Malzac}, {De Marco}, {Matt}, {Nandra}, {Paltani},
  {Peterson}, {Pinto}, {Ponti}, {Pozo Nu{\~n}ez}, {De Rosa}, {Seta}, {Ursini},
  {de Vries}, {Walton}, \& {Whewell}}]{arav2015}
{Arav}, N., {Chamberlain}, C., {Kriss}, G.~A., {et~al.} 2015, \aap, 577, A37

\bibitem[{{Beijers} {et~al.}(1992){Beijers}, {Hoekstra}, {Morgenstern}, \& {de
  Heer}}]{beijers1992}
{Beijers}, J.~P.~M., {Hoekstra}, R., {Morgenstern}, R., \& {de Heer}, F.~J.
  1992, Journal of Physics B Atomic Molecular Physics, 25, 4851

\bibitem[{{Blustin} \& {Fabian}(2009)}]{blustin2009}
{Blustin}, A.~J. \& {Fabian}, A.~C. 2009, \mnras, 399, L169

\bibitem[{{B{\"o}hm} {et~al.}(2003){B{\"o}hm}, {M{\"u}ller}, {Schippers},
  {Shi}, {Ekl{\"o}w}, {Schuch}, {Danared}, \& {Badnell}}]{bohm2003}
{B{\"o}hm}, S., {M{\"u}ller}, A., {Schippers}, S., {et~al.} 2003, \aap, 405,
  1157

\bibitem[{{Bottorff} {et~al.}(2000){Bottorff}, {Korista}, \&
  {Shlosman}}]{bottorff2000}
{Bottorff}, M.~C., {Korista}, K.~T., \& {Shlosman}, I. 2000, \apj, 537, 134

\bibitem[{{Branduardi-Raymont} {et~al.}(2001){Branduardi-Raymont}, {Sako},
  {Kahn}, {Brinkman}, {Kaastra}, \& {Page}}]{br2001}
{Branduardi-Raymont}, G., {Sako}, M., {Kahn}, S.~M., {et~al.} 2001, \aap, 365,
  L140

\bibitem[{{Cappi} {et~al.}(2016){Cappi}, {De Marco}, {Ponti}, {Ursini},
  {Petrucci}, {Bianchi}, {Kaastra}, {Kriss}, {Mehdipour}, {Whewell}, {Arav},
  {Behar}, {Boissay}, {Branduardi-Raymont}, {Costantini}, {Ebrero}, {Di Gesu},
  {Harrison}, {Kaspi}, {Matt}, {Paltani}, {Peterson}, {Steenbrugge}, \&
  {Walton}}]{cappi2016}
{Cappi}, M., {De Marco}, B., {Ponti}, G., {et~al.} 2016, \aap, 592, A27

\bibitem[{{Costantini} {et~al.}(2007){Costantini}, {Kaastra}, {Arav}, {Kriss},
  {Steenbrugge}, {Gabel}, {Verbunt}, {Behar}, {Gaskell}, {Korista}, {Proga},
  {Quijano}, {Scott}, {Klimek}, \& {Hedrick}}]{costantini2007}
{Costantini}, E., {Kaastra}, J.~S., {Arav}, N., {et~al.} 2007, \aap, 461, 121

\bibitem[{{Crenshaw} {et~al.}(2003){Crenshaw}, {Kraemer}, \&
  {George}}]{crenshaw2003}
{Crenshaw}, D.~M., {Kraemer}, S.~B., \& {George}, I.~M. 2003, \araa, 41, 117

\bibitem[{{Cumbee} {et~al.}(2018){Cumbee}, {Mullen}, {Lyons}, {Shelton},
  {Fogle}, {Schultz}, \& {Stancil}}]{cumbee2018}
{Cumbee}, R.~S., {Mullen}, P.~D., {Lyons}, D., {et~al.} 2018, \apj, 852, 7

\bibitem[{{Dauser} {et~al.}(2022){Dauser}, {Garc{\'\i}a}, {Joyce},
  {Licklederer}, {Connors}, {Ingram}, {Reynolds}, \& {Wilms}}]{dauser2022}
{Dauser}, T., {Garc{\'\i}a}, J.~A., {Joyce}, A., {et~al.} 2022, \mnras, 514,
  3965

\bibitem[{{de Vaucouleurs} {et~al.}(1991){de Vaucouleurs}, {de Vaucouleurs},
  {Corwin}, {Buta}, {Paturel}, \& {Fouque}}]{dv1991}
{de Vaucouleurs}, G., {de Vaucouleurs}, A., {Corwin}, Herold~G., J., {et~al.}
  1991, {Third Reference Catalogue of Bright Galaxies}

\bibitem[{{Dehghanian} {et~al.}(2019{\natexlab{a}}){Dehghanian}, {Ferland},
  {Kriss}, {Peterson}, {Mathur}, {Mehdipour}, {Guzm{\'a}n}, {Chatzikos}, {van
  Hoof}, {Williams}, {Arav}, {Barth}, {Bentz}, {Bisogni}, {Brandt}, {Crenshaw},
  {Dalla Bont{\`a}}, {De Rosa}, {Fausnaugh}, {Gelbord}, {Goad}, {Gupta},
  {Horne}, {Kaastra}, {Knigge}, {Korista}, {McHardy}, {Pogge}, {Starkey}, \&
  {Vestergaard}}]{deh2019}
{Dehghanian}, M., {Ferland}, G.~J., {Kriss}, G.~A., {et~al.}
  2019{\natexlab{a}}, \apj, 877, 119

\bibitem[{{Dehghanian} {et~al.}(2019{\natexlab{b}}){Dehghanian}, {Ferland},
  {Peterson}, {Kriss}, {Korista}, {Chatzikos}, {Guzm{\'a}n}, {Arav}, {De Rosa},
  {Goad}, {Mehdipour}, \& {van Hoof}}]{deh22019}
{Dehghanian}, M., {Ferland}, G.~J., {Peterson}, B.~M., {et~al.}
  2019{\natexlab{b}}, \apjl, 882, L30

\bibitem[{{Detmers} {et~al.}(2009){Detmers}, {Kaastra}, \&
  {McHardy}}]{detmers2009}
{Detmers}, R.~G., {Kaastra}, J.~S., \& {McHardy}, I.~M. 2009, \aap, 504, 409

\bibitem[{{Di Gesu} {et~al.}(2015){Di Gesu}, {Costantini}, {Ebrero},
  {Mehdipour}, {Kaastra}, {Ursini}, {Petrucci}, {Cappi}, {Kriss}, {Bianchi},
  {Branduardi-Raymont}, {De Marco}, {De Rosa}, {Kaspi}, {Paltani}, {Pinto},
  {Ponti}, {Steenbrugge}, \& {Whewell}}]{digesu2015}
{Di Gesu}, L., {Costantini}, E., {Ebrero}, J., {et~al.} 2015, \aap, 579, A42

\bibitem[{{Ebrero} {et~al.}(2016){Ebrero}, {Kaastra}, {Kriss}, {Di Gesu},
  {Costantini}, {Mehdipour}, {Bianchi}, {Cappi}, {Boissay},
  {Branduardi-Raymont}, {Petrucci}, {Ponti}, {Pozo N{\'u}{\~n}ez}, {Seta},
  {Steenbrugge}, \& {Whewell}}]{ebrero2016}
{Ebrero}, J., {Kaastra}, J.~S., {Kriss}, G.~A., {et~al.} 2016, \aap, 587, A129

\bibitem[{{Garc{\'\i}a} {et~al.}(2014){Garc{\'\i}a}, {Dauser}, {Lohfink},
  {Kallman}, {Steiner}, {McClintock}, {Brenneman}, {Wilms}, {Eikmann},
  {Reynolds}, \& {Tombesi}}]{garcia2014}
{Garc{\'\i}a}, J., {Dauser}, T., {Lohfink}, A., {et~al.} 2014, \apj, 782, 76

\bibitem[{{Giustini} {et~al.}(2017){Giustini}, {Costantini}, {De Marco},
  {Svoboda}, {Motta}, {Proga}, {Saxton}, {Ferrigno}, {Longinotti}, {Miniutti},
  {Grupe}, {Mathur}, {Shappee}, {Prieto}, \& {Stanek}}]{giustini2017}
{Giustini}, M., {Costantini}, E., {De Marco}, B., {et~al.} 2017, \aap, 597, A66

\bibitem[{{Goad} {et~al.}(2016){Goad}, {Korista}, {De Rosa}, {Kriss},
  {Edelson}, {Barth}, {Ferland}, {Kochanek}, {Netzer}, {Peterson}, {Bentz},
  {Bisogni}, {Crenshaw}, {Denney}, {Ely}, {Fausnaugh}, {Grier}, {Gupta},
  {Horne}, {Kaastra}, {Pancoast}, {Pei}, {Pogge}, {Skielboe}, {Starkey},
  {Vestergaard}, {Zu}, {Anderson}, {Ar{\'e}valo}, {Bazhaw}, {Borman},
  {Boroson}, {Bottorff}, {Brandt}, {Breeveld}, {Brewer}, {Cackett}, {Carini},
  {Croxall}, {Dalla Bont{\`a}}, {De Lorenzo-C{\'a}ceres}, {Dietrich},
  {Efimova}, {Evans}, {Filippenko}, {Flatland}, {Gehrels}, {Geier}, {Gelbord},
  {Gonzalez}, {Gorjian}, {Grupe}, {Hall}, {Hicks}, {Horenstein}, {Hutchison},
  {Im}, {Jensen}, {Joner}, {Jones}, {Kaspi}, {Kelly}, {Kennea}, {Kim}, {Kim},
  {Klimanov}, {Lee}, {Leonard}, {Lira}, {MacInnis}, {Manne-Nicholas}, {Mathur},
  {McHardy}, {Montouri}, {Musso}, {Nazarov}, {Norris}, {Nousek}, {Okhmat},
  {Papadakis}, {Parks}, {Pott}, {Rafter}, {Rix}, {Saylor}, {Schimoia},
  {Schn{\"u}lle}, {Sergeev}, {Siegel}, {Spencer}, {Sung}, {Teems}, {Treu},
  {Turner}, {Uttley}, {Villforth}, {Weiss}, {Woo}, {Yan}, {Young}, \&
  {Zheng}}]{goad2016}
{Goad}, M.~R., {Korista}, K.~T., {De Rosa}, G., {et~al.} 2016, \apj, 824, 11

\bibitem[{{Grafton-Waters} {et~al.}(2021){Grafton-Waters},
  {Branduardi-Raymont}, {Mehdipour}, {Page}, {Bianchi}, {Behar}, \&
  {Symeonidis}}]{gw2021}
{Grafton-Waters}, S., {Branduardi-Raymont}, G., {Mehdipour}, M., {et~al.} 2021,
  \aap, 649, A162

\bibitem[{{Grafton-Waters} {et~al.}(2020){Grafton-Waters},
  {Branduardi-Raymont}, {Mehdipour}, {Page}, {Behar}, {Kaastra}, {Arav},
  {Bianchi}, {Costantini}, {Ebrero}, {Di Gesu}, {Kaspi}, {Kriss}, {De Marco},
  {Mao}, {Middei}, {Peretz}, {Petrucci}, \& {Ponti}}]{gw2020}
{Grafton-Waters}, S., {Branduardi-Raymont}, G., {Mehdipour}, M., {et~al.} 2020,
  \aap, 633, A62

\bibitem[{{Gu} {et~al.}(2016){Gu}, {Kaastra}, \& {Raassen}}]{gu2016}
{Gu}, L., {Kaastra}, J., \& {Raassen}, A.~J.~J. 2016, \aap, 588, A52

\bibitem[{{Gu} {et~al.}(2015){Gu}, {Kaastra}, {Raassen}, {Mullen}, {Cumbee},
  {Lyons}, \& {Stancil}}]{gu2015}
{Gu}, L., {Kaastra}, J., {Raassen}, A.~J.~J., {et~al.} 2015, \aap, 584, L11

\bibitem[{{Gu} {et~al.}(2018){Gu}, {Mao}, {de Plaa}, {Raassen}, {Shah}, \&
  {Kaastra}}]{gu2018}
{Gu}, L., {Mao}, J., {de Plaa}, J., {et~al.} 2018, \aap, 611, A26

\bibitem[{{Gu} {et~al.}(2017){Gu}, {Mao}, {O'Dea}, {Baum}, {Mehdipour}, \&
  {Kaastra}}]{gu2017}
{Gu}, L., {Mao}, J., {O'Dea}, C.~P., {et~al.} 2017, \aap, 601, A45

\bibitem[{{Gu} {et~al.}(2019){Gu}, {Raassen}, {Mao}, {de Plaa}, {Shah},
  {Pinto}, {Werner}, {Simionescu}, {Mernier}, \& {Kaastra}}]{gu2019}
{Gu}, L., {Raassen}, A.~J.~J., {Mao}, J., {et~al.} 2019, \aap, 627, A51

\bibitem[{{Gu} {et~al.}(2020){Gu}, {Shah}, {Mao}, {Raassen}, {de Plaa},
  {Pinto}, {Akamatsu}, {Werner}, {Simionescu}, {Mernier}, {Sawada}, {Mohanty},
  {Amaro}, {Gu}, {Porter}, {Crespo L{\'o}pez-Urrutia}, \& {Kaastra}}]{gu2020}
{Gu}, L., {Shah}, C., {Mao}, J., {et~al.} 2020, \aap, 641, A93

\bibitem[{{Guainazzi} \& {Bianchi}(2007)}]{guainazzi2007}
{Guainazzi}, M. \& {Bianchi}, S. 2007, \mnras, 374, 1290

\bibitem[{{Hitomi Collaboration} {et~al.}(2018){Hitomi Collaboration},
  {Aharonian}, {Akamatsu}, {Akimoto}, {Allen}, {Angelini}, {Audard}, {Awaki},
  {Axelsson}, {Bamba}, {Bautz}, {Blandford}, {Brenneman}, {Brown}, {Bulbul},
  {Cackett}, {Chernyakova}, {Chiao}, {Coppi}, {Costantini}, {de Plaa}, {de
  Vries}, {den Herder}, {Done}, {Dotani}, {Ebisawa}, {Eckart}, {Enoto}, {Ezoe},
  {Fabian}, {Ferrigno}, {Foster}, {Fujimoto}, {Fukazawa}, {Furuzawa},
  {Galeazzi}, {Gallo}, {Gandhi}, {Giustini}, {Goldwurm}, {Gu}, {Guainazzi},
  {Haba}, {Hagino}, {Hamaguchi}, {Harrus}, {Hatsukade}, {Hayashi}, {Hayashi},
  {Hayashida}, {Hell}, {Hiraga}, {Hornschemeier}, {Hoshino}, {Hughes},
  {Ichinohe}, {Iizuka}, {Inoue}, {Inoue}, {Ishida}, {Ishikawa}, {Ishisaki},
  {Iwai}, {Kaastra}, {Kallman}, {Kamae}, {Kataoka}, {Katsuda}, {Kawai},
  {Kelley}, {Kilbourne}, {Kitaguchi}, {Kitamoto}, {Kitayama}, {Kohmura},
  {Kokubun}, {Koyama}, {Koyama}, {Kretschmar}, {Krimm}, {Kubota}, {Kunieda},
  {Laurent}, {Lee}, {Leutenegger}, {Limousin}, {Loewenstein}, {Long}, {Lumb},
  {Madejski}, {Maeda}, {Maier}, {Makishima}, {Markevitch}, {Matsumoto},
  {Matsushita}, {McCammon}, {McNamara}, {Mehdipour}, {Miller}, {Miller},
  {Mineshige}, {Mitsuda}, {Mitsuishi}, {Miyazawa}, {Mizuno}, {Mori}, {Mori},
  {Mukai}, {Murakami}, {Mushotzky}, {Nakagawa}, {Nakajima}, {Nakamori},
  {Nakashima}, {Nakazawa}, {Nobukawa}, {Nobukawa}, {Noda}, {Odaka}, {Ohashi},
  {Ohno}, {Okajima}, {Ota}, {Ozaki}, {Paerels}, {Paltani}, {Petre}, {Pinto},
  {Porter}, {Pottschmidt}, {Reynolds}, {Safi-Harb}, {Saito}, {Sakai}, {Sasaki},
  {Sato}, {Sato}, {Sato}, {Sawada}, {Schartel}, {Serlemtsos}, {Seta},
  {Shidatsu}, {Simionescu}, {Smith}, {Soong}, {Stawarz}, {Sugawara}, {Sugita},
  {Szymkowiak}, {Tajima}, {Takahashi}, {Takahashi}, {Takeda}, {Takei},
  {Tamagawa}, {Tamura}, {Tanaka}, {Tanaka}, {Tanaka}, {Tashiro}, {Tawara},
  {Terada}, {Terashima}, {Tombesi}, {Tomida}, {Tsuboi}, {Tsujimoto}, {Tsunemi},
  {Tsuru}, {Uchida}, {Uchiyama}, {Uchiyama}, {Ueda}, {Ueda}, {Uno}, {Urry},
  {Ursino}, {Watanabe}, {Werner}, {Wilkins}, {Williams}, {Yamada}, {Yamaguchi},
  {Yamaoka}, {Yamasaki}, {Yamauchi}, {Yamauchi}, {Yaqoob}, {Yatsu}, {Yonetoku},
  {Zhuravleva}, {Zoghbi}, \& {Raassen}}]{hitomiatomic}
{Hitomi Collaboration}, {Aharonian}, F., {Akamatsu}, H., {et~al.} 2018, \pasj,
  70, 12

\bibitem[{{Kaastra} \& {Bleeker}(2016)}]{kaastra2016}
{Kaastra}, J.~S. \& {Bleeker}, J.~A.~M. 2016, \aap, 587, A151

\bibitem[{{Kaastra} {et~al.}(2012){Kaastra}, {Detmers}, {Mehdipour}, {Arav},
  {Behar}, {Bianchi}, {Branduardi-Raymont}, {Cappi}, {Costantini}, {Ebrero},
  {Kriss}, {Paltani}, {Petrucci}, {Pinto}, {Ponti}, {Steenbrugge}, \& {de
  Vries}}]{kaastra2012}
{Kaastra}, J.~S., {Detmers}, R.~G., {Mehdipour}, M., {et~al.} 2012, \aap, 539,
  A117

\bibitem[{{Kaastra} {et~al.}(2014){Kaastra}, {Kriss}, {Cappi}, {Mehdipour},
  {Petrucci}, {Steenbrugge}, {Arav}, {Behar}, {Bianchi}, {Boissay},
  {Branduardi-Raymont}, {Chamberlain}, {Costantini}, {Ely}, {Ebrero}, {Di
  Gesu}, {Harrison}, {Kaspi}, {Malzac}, {De Marco}, {Matt}, {Nandra},
  {Paltani}, {Person}, {Peterson}, {Pinto}, {Ponti}, {Nu{\~n}ez}, {De Rosa},
  {Seta}, {Ursini}, {de Vries}, {Walton}, \& {Whewell}}]{kaastra2014}
{Kaastra}, J.~S., {Kriss}, G.~A., {Cappi}, M., {et~al.} 2014, Science, 345, 64

\bibitem[{{Kaastra} {et~al.}(2018){Kaastra}, {Mehdipour}, {Behar}, {Bianchi},
  {Branduardi-Raymont}, {Brenneman}, {Cappi}, {Costantini}, {De Marco}, {di
  Gesu}, {Ebrero}, {Kriss}, {Mao}, {Peretz}, {Petrucci}, {Ponti}, \&
  {Walton}}]{kaastra2018}
{Kaastra}, J.~S., {Mehdipour}, M., {Behar}, E., {et~al.} 2018, \aap, 619, A112

\bibitem[{{Kaastra} {et~al.}(2000){Kaastra}, {Mewe}, {Liedahl}, {Komossa}, \&
  {Brinkman}}]{kaastra2000}
{Kaastra}, J.~S., {Mewe}, R., {Liedahl}, D.~A., {Komossa}, S., \& {Brinkman},
  A.~C. 2000, \aap, 354, L83

\bibitem[{{Kaastra} {et~al.}(1996){Kaastra}, {Mewe}, \&
  {Nieuwenhuijzen}}]{kaastra1996}
{Kaastra}, J.~S., {Mewe}, R., \& {Nieuwenhuijzen}, H. 1996, in UV and X-ray
  Spectroscopy of Astrophysical and Laboratory Plasmas, ed. K.~{Yamashita} \&
  T.~{Watanabe}, 411--414

\bibitem[{{Kaastra} {et~al.}(2020){Kaastra}, {Raassen}, {de Plaa}, \&
  {Gu}}]{kaastra2020}
{Kaastra}, J.~S., {Raassen}, A.~J.~J., {de Plaa}, J., \& {Gu}, L. 2020, {SPEX
  X-ray spectral fitting package}

\bibitem[{{Kosec} {et~al.}(2018{\natexlab{a}}){Kosec}, {Buisson}, {Parker},
  {Pinto}, {Fabian}, \& {Walton}}]{kosec2018b}
{Kosec}, P., {Buisson}, D.~J.~K., {Parker}, M.~L., {et~al.} 2018{\natexlab{a}},
  \mnras, 481, 947

\bibitem[{{Kosec} {et~al.}(2018{\natexlab{b}}){Kosec}, {Pinto}, {Fabian}, \&
  {Walton}}]{kosec2018}
{Kosec}, P., {Pinto}, C., {Fabian}, A.~C., \& {Walton}, D.~J.
  2018{\natexlab{b}}, \mnras, 473, 5680

\bibitem[{{Kriss} {et~al.}(2019){Kriss}, {De Rosa}, {Ely}, {Peterson},
  {Kaastra}, {Mehdipour}, {Ferland}, {Dehghanian}, {Mathur}, {Edelson},
  {Korista}, {Arav}, {Barth}, {Bentz}, {Brandt}, {Crenshaw}, {Dalla Bont{\`a}},
  {Denney}, {Done}, {Eracleous}, {Fausnaugh}, {Gardner}, {Goad}, {Grier},
  {Horne}, {Kochanek}, {McHardy}, {Netzer}, {Pancoast}, {Pei}, {Pogge},
  {Proga}, {Silva}, {Tejos}, {Vestergaard}, {Adams}, {Anderson}, {Ar{\'e}valo},
  {Beatty}, {Behar}, {Bennert}, {Bianchi}, {Bigley}, {Bisogni},
  {Boissay-Malaquin}, {Borman}, {Bottorff}, {Breeveld}, {Brotherton}, {Brown},
  {Brown}, {Cackett}, {Canalizo}, {Cappi}, {Carini}, {Clubb}, {Comerford},
  {Coker}, {Corsini}, {Costantini}, {Croft}, {Croxall}, {Deason}, {De
  Lorenzo-C{\'a}ceres}, {De Marco}, {Dietrich}, {Di Gesu}, {Ebrero}, {Evans},
  {Filippenko}, {Flatland}, {Gates}, {Gehrels}, {Geier}, {Gelbord}, {Gonzalez},
  {Gorjian}, {Grupe}, {Gupta}, {Hall}, {Henderson}, {Hicks}, {Holmbeck},
  {Holoien}, {Hutchison}, {Im}, {Jensen}, {Johnson}, {Joner}, {Kaspi}, {Kelly},
  {Kelly}, {Kennea}, {Kim}, {Kim}, {Kim}, {King}, {Klimanov}, {Krongold},
  {Lau}, {Lee}, {Leonard}, {Li}, {Lira}, {Lochhaas}, {Ma}, {MacInnis},
  {Malkan}, {Manne-Nicholas}, {Matt}, {Mauerhan}, {McGurk}, {Montuori},
  {Morelli}, {Mosquera}, {Mudd}, {M{\"u}ller-S{\'a}nchez}, {Nazarov}, {Norris},
  {Nousek}, {Nguyen}, {Ochner}, {Okhmat}, {Paltani}, {Parks}, {Pinto},
  {Pizzella}, {Poleski}, {Ponti}, {Pott}, {Rafter}, {Rix}, {Runnoe}, {Saylor},
  {Schimoia}, {Schn{\"u}lle}, {Scott}, {Sergeev}, {Shappee}, {Shivvers},
  {Siegel}, {Simonian}, {Siviero}, {Skielboe}, {Somers}, {Spencer}, {Starkey},
  {Stevens}, {Sung}, {Tayar}, {Teems}, {Treu}, {Turner}, {Uttley}, {. Van
  Saders}, {Vican}, {Villforth}, {Villanueva}, {Walton}, {Waters}, {Weiss},
  {Woo}, {Yan}, {Yuk}, {Zheng}, {Zhu}, \& {Zu}}]{kriss2019}
{Kriss}, G.~A., {De Rosa}, G., {Ely}, J., {et~al.} 2019, \apj, 881, 153

\bibitem[{{Landt} {et~al.}(2019){Landt}, {Ward}, {Kynoch}, {Packham},
  {Ferland}, {Lawrence}, {Pott}, {Esser}, {Horne}, {Starkey}, {Malhotra},
  {Fausnaugh}, {Peterson}, {Wilman}, {Riffel}, {Storchi-Bergmann}, {Barth},
  {Villforth}, \& {Winkler}}]{landt2019}
{Landt}, H., {Ward}, M.~J., {Kynoch}, D., {et~al.} 2019, \mnras, 489, 1572

\bibitem[{{Landt} {et~al.}(2015){Landt}, {Ward}, {Steenbrugge}, \&
  {Ferland}}]{landt2015}
{Landt}, H., {Ward}, M.~J., {Steenbrugge}, K.~C., \& {Ferland}, G.~J. 2015,
  \mnras, 454, 3688

\bibitem[{{Laor}(1991)}]{laor1991}
{Laor}, A. 1991, \apj, 376, 90

\bibitem[{{Lee} {et~al.}(1991){Lee}, {Richard}, {Sanders}, {Zouros},
  {Shinpaugh}, \& {Varghese}}]{lee1991}
{Lee}, D.~H., {Richard}, P., {Sanders}, J.~M., {et~al.} 1991, \pra, 44, 1636

\bibitem[{{Lee}(2010)}]{julia2010}
{Lee}, J.~C. 2010, \ssr, 157, 93

\bibitem[{{Lee} {et~al.}(2001){Lee}, {Ogle}, {Canizares}, {Marshall}, {Schulz},
  {Morales}, {Fabian}, \& {Iwasawa}}]{julia2001}
{Lee}, J.~C., {Ogle}, P.~M., {Canizares}, C.~R., {et~al.} 2001, \apjl, 554, L13

\bibitem[{{Li} {et~al.}(2016){Li}, {Wang}, {Ho}, {Lu}, {Qiu}, {Du}, {Hu},
  {Huang}, {Zhang}, {Wang}, \& {Bai}}]{li2016}
{Li}, Y.-R., {Wang}, J.-M., {Ho}, L.~C., {et~al.} 2016, \apj, 822, 4

\bibitem[{{Longinotti} {et~al.}(2010){Longinotti}, {Costantini}, {Petrucci},
  {Boisson}, {Mouchet}, {Santos-Lleo}, {Matt}, {Ponti}, \&
  {Gon{\c{c}}alves}}]{long2010}
{Longinotti}, A.~L., {Costantini}, E., {Petrucci}, P.~O., {et~al.} 2010, \aap,
  510, A92

\bibitem[{{Mao} {et~al.}(2018){Mao}, {Kaastra}, {Mehdipour}, {Gu},
  {Costantini}, {Kriss}, {Bianchi}, {Branduardi-Raymont}, {Behar}, {Di Gesu},
  {Ponti}, {Petrucci}, \& {Ebrero}}]{mao2018}
{Mao}, J., {Kaastra}, J.~S., {Mehdipour}, M., {et~al.} 2018, \aap, 612, A18

\bibitem[{{Mao} {et~al.}(2017){Mao}, {Kaastra}, {Mehdipour}, {Raassen}, {Gu},
  \& {Miller}}]{mao2017}
{Mao}, J., {Kaastra}, J.~S., {Mehdipour}, M., {et~al.} 2017, \aap, 607, A100

\bibitem[{{Mao} {et~al.}(2019){Mao}, {Mehdipour}, {Kaastra}, {Costantini},
  {Pinto}, {Branduardi-Raymont}, {Behar}, {Peretz}, {Bianchi}, {Kriss},
  {Ponti}, {De Marco}, {Petrucci}, {Di Gesu}, {Middei}, {Ebrero}, \&
  {Arav}}]{mao2019}
{Mao}, J., {Mehdipour}, M., {Kaastra}, J.~S., {et~al.} 2019, \aap, 621, A99

\bibitem[{{McKernan} {et~al.}(2007){McKernan}, {Yaqoob}, \&
  {Reynolds}}]{mckernan2007}
{McKernan}, B., {Yaqoob}, T., \& {Reynolds}, C.~S. 2007, \mnras, 379, 1359

\bibitem[{{Mehdipour} {et~al.}(2017){Mehdipour}, {Kaastra}, {Kriss}, {Arav},
  {Behar}, {Bianchi}, {Branduardi-Raymont}, {Cappi}, {Costantini}, {Ebrero},
  {Di Gesu}, {Kaspi}, {Mao}, {De Marco}, {Matt}, {Paltani}, {Peretz},
  {Peterson}, {Petrucci}, {Pinto}, {Ponti}, {Ursini}, {de Vries}, \&
  {Walton}}]{mehd2017}
{Mehdipour}, M., {Kaastra}, J.~S., {Kriss}, G.~A., {et~al.} 2017, \aap, 607,
  A28

\bibitem[{{Mehdipour} {et~al.}(2015){Mehdipour}, {Kaastra}, {Kriss}, {Cappi},
  {Petrucci}, {Steenbrugge}, {Arav}, {Behar}, {Bianchi}, {Boissay},
  {Branduardi-Raymont}, {Costantini}, {Ebrero}, {Di Gesu}, {Harrison}, {Kaspi},
  {De Marco}, {Matt}, {Paltani}, {Peterson}, {Ponti}, {Pozo Nu{\~n}ez}, {De
  Rosa}, {Ursini}, {de Vries}, {Walton}, \& {Whewell}}]{mehd2015}
{Mehdipour}, M., {Kaastra}, J.~S., {Kriss}, G.~A., {et~al.} 2015, \aap, 575,
  A22

\bibitem[{{Mehdipour} {et~al.}(2022){Mehdipour}, {Kriss}, {Brenneman},
  {Costantini}, {Kaastra}, {Branduardi-Raymont}, {Di Gesu}, {Ebrero}, \&
  {Mao}}]{mehd2022}
{Mehdipour}, M., {Kriss}, G.~A., {Brenneman}, L.~W., {et~al.} 2022, \apj, 925,
  84

\bibitem[{{Mullen} {et~al.}(2016){Mullen}, {Cumbee}, {Lyons}, \&
  {Stancil}}]{mullen2016}
{Mullen}, P.~D., {Cumbee}, R.~S., {Lyons}, D., \& {Stancil}, P.~C. 2016, \apjs,
  224, 31

\bibitem[{{Nolte} {et~al.}(2012){Nolte}, {Stancil}, {Liebermann}, {Buenker},
  {Hui}, \& {Schultz}}]{nolte2012}
{Nolte}, J.~L., {Stancil}, P.~C., {Liebermann}, H.~P., {et~al.} 2012, Journal
  of Physics B Atomic Molecular Physics, 45, 245202

\bibitem[{{Parker} {et~al.}(2018){Parker}, {Buisson}, {Jiang}, {Gallo}, {Kara},
  {Matzeu}, \& {Walton}}]{parker2018}
{Parker}, M.~L., {Buisson}, D.~J.~K., {Jiang}, J., {et~al.} 2018, \mnras, 479,
  L45

\bibitem[{{Parker} {et~al.}(2017){Parker}, {Pinto}, {Fabian}, {Lohfink},
  {Buisson}, {Alston}, {Kara}, {Cackett}, {Chiang}, {Dauser}, {De Marco},
  {Gallo}, {Garcia}, {Harrison}, {King}, {Middleton}, {Miller}, {Miniutti},
  {Reynolds}, {Uttley}, {Vasudevan}, {Walton}, {Wilkins}, \&
  {Zoghbi}}]{parker2017}
{Parker}, M.~L., {Pinto}, C., {Fabian}, A.~C., {et~al.} 2017, \nat, 543, 83

\bibitem[{{Pepmiller} {et~al.}(1983){Pepmiller}, {Richard}, {Newcomb},
  {Dillingham}, {Hall}, {Gray}, \& {Stockli}}]{pepmiller1983}
{Pepmiller}, P.~L., {Richard}, P., {Newcomb}, J., {et~al.} 1983, IEEE
  Transactions on Nuclear Science, 30, 1002

\bibitem[{{Pinto} {et~al.}(2018){Pinto}, {Alston}, {Parker}, {Fabian}, {Gallo},
  {Buisson}, {Walton}, {Kara}, {Jiang}, {Lohfink}, \& {Reynolds}}]{pinto2018}
{Pinto}, C., {Alston}, W., {Parker}, M.~L., {et~al.} 2018, \mnras, 476, 1021

\bibitem[{{Pinto} {et~al.}(2021){Pinto}, {Soria}, {Walton}, {D'A{\`\i}},
  {Pintore}, {Kosec}, {Alston}, {Fuerst}, {Middleton}, {Roberts}, {Del Santo},
  {Barret}, {Ambrosi}, {Robba}, {Earnshaw}, \& {Fabian}}]{pinto2021}
{Pinto}, C., {Soria}, R., {Walton}, D.~J., {et~al.} 2021, \mnras, 505, 5058

\bibitem[{{Pinto} {et~al.}(2020){Pinto}, {Walton}, {Kara}, {Parker}, {Soria},
  {Kosec}, {Middleton}, {Alston}, {Fabian}, {Guainazzi}, {Roberts}, {Fuerst},
  {Earnshaw}, {Sathyaprakash}, \& {Barret}}]{pinto2020}
{Pinto}, C., {Walton}, D.~J., {Kara}, E., {et~al.} 2020, \mnras, 492, 4646

\bibitem[{{Pounds} {et~al.}(2018){Pounds}, {Nixon}, {Lobban}, \&
  {King}}]{pounds2018}
{Pounds}, K.~A., {Nixon}, C.~J., {Lobban}, A., \& {King}, A.~R. 2018, \mnras,
  481, 1832

\bibitem[{{Raphaelian} {et~al.}(1991){Raphaelian}, {Berry}, {Berrah Mansour},
  \& {Schneider}}]{raphaelian1991}
{Raphaelian}, M.~L.~A., {Berry}, H.~G., {Berrah Mansour}, N., \& {Schneider},
  D. 1991, \pra, 43, 4071

\bibitem[{{Reeves} {et~al.}(2020){Reeves}, {Braito}, {Chartas}, {Hamann},
  {Laha}, \& {Nardini}}]{reeves2020}
{Reeves}, J.~N., {Braito}, V., {Chartas}, G., {et~al.} 2020, \apj, 895, 37

\bibitem[{{Reynolds}(2016)}]{reynolds2016}
{Reynolds}, C.~S. 2016, Astronomische Nachrichten, 337, 404

\bibitem[{{Shapovalova} {et~al.}(2004){Shapovalova}, {Doroshenko}, {Bochkarev},
  {Burenkov}, {Carrasco}, {Chavushyan}, {Collin}, {Vald{\'e}s}, {Borisov},
  {Dumont}, {Vlasuyk}, {Chilingarian}, {Fioktistova}, \&
  {Martinez}}]{shapovalova2004}
{Shapovalova}, A.~I., {Doroshenko}, V.~T., {Bochkarev}, N.~G., {et~al.} 2004,
  \aap, 422, 925

\bibitem[{{Smith} {et~al.}(2012){Smith}, {Foster}, \& {Brickhouse}}]{smith2012}
{Smith}, R.~K., {Foster}, A.~R., \& {Brickhouse}, N.~S. 2012, Astronomische
  Nachrichten, 333, 301

\bibitem[{{Smith} {et~al.}(2014){Smith}, {Foster}, {Edgar}, \&
  {Brickhouse}}]{smith2014}
{Smith}, R.~K., {Foster}, A.~R., {Edgar}, R.~J., \& {Brickhouse}, N.~S. 2014,
  \apj, 787, 77

\bibitem[{{Steenbrugge} {et~al.}(2005){Steenbrugge}, {Kaastra}, {Crenshaw},
  {Kraemer}, {Arav}, {George}, {Liedahl}, {van der Meer}, {Paerels}, {Turner},
  \& {Yaqoob}}]{steenbrugge2005}
{Steenbrugge}, K.~C., {Kaastra}, J.~S., {Crenshaw}, D.~M., {et~al.} 2005, \aap,
  434, 569

\bibitem[{{Tanis} {et~al.}(1985){Tanis}, {Bernstein}, {Clark}, {Graham},
  {McFarland}, {Morgan}, {Johnson}, {Jones}, \& {Meron}}]{tanis1985}
{Tanis}, J.~A., {Bernstein}, E.~M., {Clark}, M.~W., {et~al.} 1985, \pra, 31,
  4040

\bibitem[{{Tombesi} {et~al.}(2013){Tombesi}, {Cappi}, {Reeves}, {Nemmen},
  {Braito}, {Gaspari}, \& {Reynolds}}]{tombesi2013}
{Tombesi}, F., {Cappi}, M., {Reeves}, J.~N., {et~al.} 2013, \mnras, 430, 1102

\bibitem[{{Tombesi} {et~al.}(2010){Tombesi}, {Cappi}, {Reeves}, {Palumbo},
  {Yaqoob}, {Braito}, \& {Dadina}}]{tombesi2010}
{Tombesi}, F., {Cappi}, M., {Reeves}, J.~N., {et~al.} 2010, \aap, 521, A57

\bibitem[{{Turner} \& {Miller}(2009)}]{turner2009}
{Turner}, T.~J. \& {Miller}, L. 2009, \aapr, 17, 47

\bibitem[{{Ursini} {et~al.}(2015){Ursini}, {Boissay}, {Petrucci}, {Matt},
  {Cappi}, {Bianchi}, {Kaastra}, {Harrison}, {Walton}, {di Gesu}, {Costantini},
  {De Marco}, {Kriss}, {Mehdipour}, {Paltani}, {Peterson}, {Ponti}, \&
  {Steenbrugge}}]{ursini2015}
{Ursini}, F., {Boissay}, R., {Petrucci}, P.~O., {et~al.} 2015, \aap, 577, A38

\bibitem[{{Vaughan} \& {Uttley}(2008)}]{vaughan2008}
{Vaughan}, S. \& {Uttley}, P. 2008, \mnras, 390, 421

\bibitem[{{Wakker} {et~al.}(2011){Wakker}, {Lockman}, \& {Brown}}]{wakker2011}
{Wakker}, B.~P., {Lockman}, F.~J., \& {Brown}, J.~M. 2011, \apj, 728, 159

\bibitem[{{Whewell} {et~al.}(2015){Whewell}, {Branduardi-Raymont}, {Kaastra},
  {Mehdipour}, {Steenbrugge}, {Bianchi}, {Behar}, {Ebrero}, {Cappi},
  {Costantini}, {De Marco}, {Di Gesu}, {Kriss}, {Paltani}, {Peterson},
  {Petrucci}, {Pinto}, \& {Ponti}}]{whewell2015}
{Whewell}, M., {Branduardi-Raymont}, G., {Kaastra}, J.~S., {et~al.} 2015, \aap,
  581, A79

\bibitem[{{Wildy} {et~al.}(2021){Wildy}, {Landt}, {Ward}, {Czerny}, \&
  {Kynoch}}]{wildy2021}
{Wildy}, C., {Landt}, H., {Ward}, M.~J., {Czerny}, B., \& {Kynoch}, D. 2021,
  \mnras, 500, 2063

\bibitem[{{Wu} {et~al.}(2012){Wu}, {Stancil}, {Schultz}, {Hui}, {Liebermann},
  \& {Buenker}}]{wu2012}
{Wu}, Y., {Stancil}, P.~C., {Schultz}, D.~R., {et~al.} 2012, Journal of Physics
  B Atomic Molecular Physics, 45, 235201

\bibitem[{{Xu} {et~al.}(2021){Xu}, {Pinto}, {Bianchi}, {Kosec}, {Parker},
  {Walton}, {Fabian}, {Guainazzi}, {Barret}, \& {Cusumano}}]{xu2021}
{Xu}, Y., {Pinto}, C., {Bianchi}, S., {et~al.} 2021, \mnras, 508, 6049

\end{thebibliography}

\begin{appendix}
\section{Significance of the excess}
\label{app}

The $\Delta$C-stat reported in \S~\ref{detection} cannot be directly used to determine the confidence level of the feature. This is due to the parameter space and wavelength range explored in the Gaussian scan, and the possibility of detecting a random excess feature from the look-elsewhere effect.

First we address this effect in an analytic way. Assuming that the residual data points ($Z_{i}, i = [1,n]$) follow a normal distribution, we define the probability of an individual data point $P(Z_{i} \leq X) = 1 - \Phi(X)$, where $X$ is the confidence level, and $\Phi(X)$ is the likelihood that the null hypothesis is true. To have all $Z_{i} \leq X$, the probability becomes $P(Z_{i} \leq X)^{n} \approx 1 - n \Phi(X)$ for $\Phi(X) \ll 1$, where $n$ is the number of effective resolving units. 

The size of resolving unit can be obtained as $\delta \lambda = (\delta v / c) \lambda $, where $\delta v = 2.35 \sigma$ is the full width at half maximum of the scanning Gaussian component, and $c$ is the speed of light. The total number of resolving units in the $5-40$~{\AA} band is 

\begin{equation}
n = \int_{5}^{40} \frac{d\lambda}{\delta \lambda} = \frac{c}{\delta v} \int_{5}^{40} \frac{d\lambda}{\lambda} 
= \frac{2.66 \times 10^{5} \;\; \rm (km \; s^{-1}) }{\sigma}. 
\end{equation}

For the Gaussian $\sigma = 1000$ km s$^{-1}$ and 4000 km s$^{-1}$, the resolving unit numbers are 266 and 67 per instrument. This numbers are determined independent of the instruments used. For the observed C-stat improvements $\Delta$C-stat $=$ 57 and 71 (\S~\ref{detection}), the null hypothesis probabilities $\Phi(X)$ are $2.2 \times 10^{-14}$ and $1.8 \times 10^{-17}$ at each individual data point, respectively. The modified probabilities $n \Phi(X)$ taking into account the look-elsewhere effect of the scan become $1.2 \times 10^{-11}$ and $2.4 \times 10^{-15}$. Therefore, the significances of the feature are estimated to be 6.6$\sigma$ and 7.8$\sigma$ for the two kinds of Gaussian widths considered.

\begin{figure}[!htbp]
\centering
\resizebox{1.0\hsize}{!}{\includegraphics[angle=0]{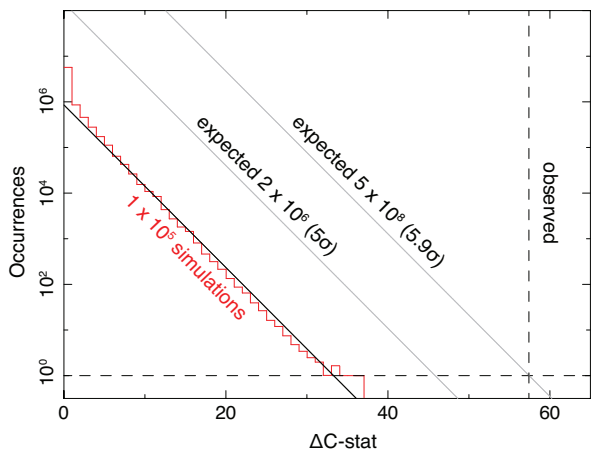}}
\caption{Histogram of $1\times 10^{5}$ runs of Monto Carlo simulation of the NGC~5548 spectrum. The thick black line represents
the corresponding power-law fit of the histogram, and the thin black lines show the expected results of $2 \times 10^{6}$
and $5 \times 10^{8}$ simulations. The vertical dashed line marks the observed $\Delta$C-stat = 57 derived with the scan on the observed data using a Gaussian line with $\sigma = $ 1000 km s$^{-1}$. 
}
\label{fig:lookelse}
\end{figure}


Besides the analytic approach, it is also possible to quantify the probability of detecting random features using a Monte-Carlo simulation. Following the approach of \citet{kosec2018} and \citet{pinto2020}, we simulate $1\times 10^{5}$ LETGS and RGS spectra based on the baseline model. The LETGS and RGS exposures are set to the real values. The residual of each set of simulated spectra
has been scanned using the Gaussian line with $\sigma = 1000$ km s$^{-1}$, the occurrence of $\Delta$C-stat improvement has been recorded in Figure~\ref{fig:lookelse}. The logarithm of the occurrence can be well described by a linear equation with a slope of $-0.178 \pm 0.007$, which allows to infer a $\Delta$C-stat $= 33.3 \pm 1.2$ for a 4.4$\sigma$ event with an expected frequency of $1\times 10^{-5}$. 

It is too computationally expensive to calculate directly the $\Delta$C-stat distribution for 5$\sigma$ which requires a sample of $2 \times 10^6$. However, it would be possible to extrapolate from the existing $1\times 10^{5}$ run. As reported in \citet{pinto2021}, the slope of $\Delta$C-stat distribution obtained with $2 \times 10^{4}$ simulation on a grating spectrum appears to be in good agreement with those obtained with $2 \times 10^{3}$ and $5 \times 10^{4}$ simulations. Therefore we predict the $\Delta$C-stat distribution of $2 \times 10^{6}$ simulations using the linear equation with slope of $-0.178$. It would suggest a 5$\sigma$ detection with a $\Delta$C-stat $= 45.9 \pm 1.6$. 

As shown in Figure~\ref{fig:lookelse}, the above scaling predicts that the observed total 
$\Delta$C-stat $= 57$ is consistent with a confidence level of about 5.9$\sigma$. This is slightly lower than the significance (6.6$\sigma$) obtained in the analytic way. Since the numerical simulations produce more realistic distributions of the spectral residuals, we consider the numerical value (5.9$\sigma$) as a better estimate of the confidence level of the putative 18.4~{\AA} feature.

\end{appendix}

\end{document}